\documentclass{jfm}

\usepackage{newtxtext}
\usepackage{newtxmath}
\usepackage{natbib}
\usepackage{hyperref}
\hypersetup{
    colorlinks = true,
    urlcolor   = blue,
    citecolor  = black,
}
\usepackage{graphicx}
\usepackage{tikz}
\usepackage{epstopdf, epsfig}
\usepackage{dirtytalk}
\usepackage{amssymb}
\usepackage{amsmath}
\DeclareMathOperator{\sgn}{sgn}
\usepackage{comment}
\usepackage{caption}
\usepackage{listings,amsmath,amssymb,xcolor,graphicx,url,natbib}
\usepackage{tikz}
\usepackage{pgfplots}
\pgfplotsset{compat=1.17}
\usepgfplotslibrary{fillbetween}
\usetikzlibrary{intersections}
\usepackage{subfig}

\newcommand{\RomanNumeralCaps}[1]

\shorttitle{Evolution of a viscoplastic liquid coating a cylindrical tube}
\shortauthor{J. D. Shemilt, A. Horsley, O. E. Jensen, A. B. Thompson and C. A. Whitfield}

\title{Surface-tension-driven evolution of a viscoplastic liquid coating the interior of a cylindrical tube
}

\author{James D. Shemilt\aff{1}
  \corresp{\email{james.shemilt@manchester.ac.uk}},
  Alexander Horsley\aff{2}, Oliver E. Jensen\aff{1}, Alice B. Thompson\aff{1}
 \and Carl A. Whitfield\aff{1,2}}

\affiliation{\aff{1}Department of Mathematics, University of Manchester,
Manchester, M13 9PL, UK
\aff{2}Division of Immunology, Immunity to Infection and Respiratory Medicine, University of
Manchester, Manchester, M13 9PL, UK}

\begin{document}
\maketitle

\begin{abstract}
One mechanism for airway closure in the lung is the surface-tension-driven instability of the mucus layer which lines the airway wall. We study the instability of an axisymmetric layer of viscoplastic Bingham liquid coating the interior of a rigid tube, which is a simple model for an airway that takes into account the yield stress of mucus. An evolution equation for the thickness of the liquid layer is derived using long-wave theory, from which we also derive a simpler thin-film evolution equation. In the thin-film case, we show that two branches of marginally-yielded static solutions of the evolution equation can be used to both predict the size of initial perturbation required to trigger instability and quantify how increasing the capillary Bingham number (a parameter measuring yield stress relative to surface tension) reduces the final deformation of the layer. Using numerical solutions of the long-wave evolution equation, we quantify how the critical layer thickness required to form a liquid plug in the tube increases as the capillary Bingham number is increased. We discuss the significance of these findings for modelling airway closure in obstructive conditions such as cystic fibrosis, where the mucus layer is often thicker and has a higher yield stress.
\end{abstract}

\section{Introduction}

The surface-tension-driven instability of a liquid layer lining the interior of a cylindrical tube is of physiological importance since, when it occurs in a lung airway, it can cause obstruction or airway closure by redistributing the liquid lining the airway, potentially leading to formation of a liquid plug. The liquid that lines the lung's airways consists primarily of mucus, a non-Newtonian fluid exhibiting various rheological properties such as shear-thinning and viscoelasticity \citep{hill_2022_mucus}. Importantly, mucus also has a yield stress, which is significantly increased in diseases such as cystic fibrosis (CF) and chronic obstructive pulmonary disease (COPD) compared to its typical value in healthy lungs \citep{patarin_rheological_2020}. Increased prevalence of airway obstruction by mucus plugging is also a key symptom of CF and COPD \citep{mall_unplugging_2016}. Motivated by this application, we study the effect of viscoplastic liquid rheology on the evolution of a layer coating the interior of a cylindrical tube, which is a simple model for an airway. Additionally, there are numerous potential applications of this class of flow in engineering and industry, as highlighted by \citet{craster_dynamics_2009} for thin-film and coating flows, and \citet{balmforth_yielding_2014} for thin-film and free-surface viscoplastic flows. 

The surface-tension-driven flow of a viscous film coating a rigid circular cylinder has been well studied in the case that the liquid layer is Newtonian. \citet{Goren_1962} identified that a flat layer can be linearly unstable. \citet{everett_model_1972} found and analysed capillary-static configurations of a volume of liquid inside a tube, which were either annular collars of fluid or liquid plugs. A nonlinear evolution equation using thin-film theory was first derived and solved by \citet{hammond_nonlinear_1983}, who found that an initially flat layer evolves into a configuration with large quasi-static annular collars separated by thin films which slowly drain into the collars. \citet{lister_capillary_2006} studied the long-time dynamics of the thin-film system, finding that at very long times collars can translate along the tube, potentially consuming other collars in the process, provided other physical effects do not intervene first, while \citet{xu_trapping_2017} showed how collars can be pinned by wall roughness.
Hammond's theory was extended by \citet{gauglitz_extended_1988} to predict plug formation by retaining certain higher-order terms in the thin-film theory, notably the exact free-surface curvature. This approach provides a composite approximation to the evolution of layers with thickness comparable to the tube radius, which accurately determines capillary-static effects whilst approximating the dynamics well where the layer is thin. They identified a critical average layer thickness, approximately $12\%$ of the tube radius, required for a liquid plug to form during the evolution. A similar composite approximation was compared to full two-dimensional numerical simulations by \citet{johnson_nonlinear_1991}; whilst their quasi-one-dimensional theory could predict when a plug would form, it could not capture the genuinely two-dimensional dynamics which occur around coalescence. \citet{otis_role_1993} derived a similar reduced-order model by making a long-wave assumption when simplifying the governing equations. 

At this point, we clarify the distinction between thin-film and long-wave approaches to deriving reduced-order evolution equations: in thin-film theory, it is assumed that the thickness of the layer is much smaller than the radius of the tube, while in long-wave theory, it is assumed that the tube radius is much smaller than the characteristic axial lengthscale of the flow but the layer is not necessarily thin compared to the radius. Making the thin-film assumption results in an evolution equation with the same mobility function as would appear in the planar case, and the curvature of the cylindrical geometry is felt only through the linearised free-surface curvature. In long-wave theory, additional terms appear in the mobility which arise due to the curvature of the geometry, and the full expression for the free-surface curvature is generally retained. Thin-film models benefit from their relative simplicity, but long-wave models capture the effects of the curved geometry more accurately \citep{camassa_viscous_2015}, allowing the dynamics leading to plug formation to be described. 

Various physical effects that can modify the evolution of the coating layer have previously been incorporated into models. \citet{halpern_effect_2010} used a long-wave evolution equation to model the effect of viscoelasticity, showing that the critical layer thickness for plug formation is not changed but the time to form a plug can be shortened by increasing the Weissenberg number (a parameter proportional to the relaxation time of the fluid). \citet{romano_liquid_2019} used a volume-of-fluid method to model the pre- and post-coalescence phases of Newtonian plug formation, and recently extended this work to include the effect of viscoelasticity \citep{romano_effect_2021}. They found that post-coalescence bi-frontal plug growth can induce significant stresses on the tube wall, and that viscoelasticity can induce additional wall stress due to the occurrence of an elastic instability. \citet{Erken_capillary_2022} studied the instability of a two-layer coating on the interior of a tube as a model for a lung airway which takes into account the periciliary liquid layer that lies beneath the mucus layer and is generally less viscous than mucus. They found that plug formation occurs more quickly in a two-layer model due to the lubricating effect of the base layer, but that the combined critical thickness of the layers required for plug formation can be significantly larger than the single-layer result of \citet{gauglitz_extended_1988}. 
\citet{halpern_fluid-elastic_1992} modelled the evolution of a liquid layer coating an elastic tube and subsequently extended the model to include the effect of insoluble surfactants \citep{halpern_surfactant_1993}. They found that the presence of surfactant can significantly increase the critical layer thickness required to form a plug and can delay plug formation when it does occur, while decreasing the wall stiffness has the opposite effect, decreasing both the critical layer thickness for plug formation and the closure time. \citet{heil_mechanics_2008} showed that the volume of liquid required for a plug to form is significantly decreased if there is non-axisymmetric collapse of the elastic tube wall. \citet{halpern_nonlinear_2003} developed a thin-film model that included the effect of an oscillating air-flow in the centre of the tube, showing that at certain frequencies of oscillation, air-flow can suppress deformation of the liquid layer. \citet{camassa_2014_gravity} developed a long-wave model for gravity-driven flow, and identified families of travelling-wave solutions which they found can be used to predict the critical thickness for plug formation as a function of the Bond number. \citet{camassa_2017_air} also found travelling-wave solutions for the case of flow driven by air-flow in the centre of the tube and recently \citet{ogrosky_linear_2021} extended the long-wave model to include the combined effects of gravity, air-flow and surfactant. 

Turning to viscoplastic flows with applications in airway modelling, \citet{craster_surfactant_2000} modelled surfactant-driven flow on a single-layer or two-layer film of viscoplastic or Newtonian fluids, showing that, at least in the single-layer case, yield stress decreases spreading rates and can cause the layer to become frozen in a non-trivial static shape. Modelling of propagation of viscoplastic liquid plugs in tubes and channels has shown that increasing the yield stress increases the stress applied to the wall and increases the
thickness of the layer of liquid left behind as a plug propagates \citep{zamankhan_steady_2012,zamankhan_steady_2018}. Rupture of viscoplastic liquid plugs has also been modelled both experimentally \citep{hu_microfluidic_2015} and numerically \citep{hu_effects_2020}, showing that increased yield stress can inhibit plug rupture because a larger pressure drop is required across the plug to make it yield. Recently, \citet{Bahrani_propagation_2022} proposed a model of elastoviscoplastic plugs, which they validated against experimental results, showing that increased yield stress slows the propagation of a plug but can speed up its rupture since the trailing film thickness is increased. The distribution of mucus throughout a whole lung has also been studied using a viscoplastic model for mucus, showing that the yield stress and the strength of air-flow (in this case modelling air-flow induced by chest physiotherapy) influence the mucus layer thickness in each airway generation \citep{mauroy_toward_2011,mauroy_toward_2015}.

The exposition of viscoplastic thin-film theory by \citet{BALMFORTH199965} has provided the basis for various studies of canonical viscoplastic free-surface flows. In their theory, there are regions of plug-like flow near the free-surface, which have the same structure as the \say{pseudo-plugs} first identified in a bounded annular flow by \citet{walton_axial_1991}.
Viscoplastic thin-film theory was used by \citet{balmforth_visco-plastic_2000} to study axisymmetrically spreading gravity currents and the work was recently extended to model droplets spreading under surface tension as well as gravity \citep{jalaal_stoeber_balmforth_2021}, showing that after spreading, the fluid is frozen in a non-trivial static shape in which the hydrostatic or capillary pressure is balanced by resistance from the yield stress. \citet{jalaal_stoeber_balmforth_2021} also compared results for the final shape and size of the droplets computed using thin-film theory to results from computational fluid dynamics (CFD) simulations showing good agreement except when the capillary Bingham number was very large. Gravity-driven flow down inclined planes has been well studied using viscoplastic thin-film theory, as reviewed by \citet{BALMFORTH2007219}. \citet{balmforth_surface_2007} investigated the surface-tension-driven fingering instability of a film travelling down an inclined plane, finding that increasing the Bingham number (which measures yield stress relative to viscous stress) slows growth of the linear instability and, when it is above a critical value, instability is fully suppressed. \citet{jalaal_long_2016} also used thin-film theory to model the steady propagation of a bubble through a tube filled with viscoplastic fluid, and compared their results to CFD simulations, showing that thin-film theory predicts the behaviour accurately when the liquid film is thin but less well when the Bingham number is increased and the film is thicker. Viscoplastic flows are often solved using regularisation of the constitutive equation; \citet{frigaard_usage_2005} review popular regularisation approaches. \citet{jalaal_thesis_2016} introduced a regularisation specifically designed for thin-film flows which we describe in §\ref{section:solutionmethods} and use when solving our evolution equations numerically.

With this study, we aim to quantify the effect of viscoplastic liquid rheology on the surface-tension-driven Rayleigh-Plateau instability of a layer coating the interior of a cylindrical tube. This flow has not previously been studied in the case that the liquid layer is viscoplastic. We ask how the yield stress affects the dynamics during the evolution of the layer and the critical layer thickness required to form a plug. To answer these questions, we derive an evolution equation using long-wave theory, with the detailed flow structure inspired by the viscoplastic thin-film theory of \citet{BALMFORTH199965}, but we include additional terms arising from the cylindrical geometry which are neglected in the thin-film approximation. We then show how this model reduces, in the appropriate limit, to a thin-film evolution equation, analogous to the Newtonian version derived by \citet{hammond_nonlinear_1983}. \textcolor{black}{Other complicating effects are neglected in the model so that the effect of the viscoplastic rheology can be examined in isolation: the tube is rigid, the flow is axisymmetric, surface tension is constant, and the air in the centre of the tube is passive and inviscid. We use the Bingham model for the liquid layer since it is the simplest viscoplastic rheology, without the potentially complicating effects of, for example, shear-thinning, elasticity or thixotropy.} We compute marginally-yielded static solutions to the thin-film evolution equation, and show how they can be used to predict both the size of perturbation to a flat layer required to trigger instability, and the final shape of the layer when there is instability. The thin-film theory cannot, however, predict formation of liquid plugs. By solving the long-wave evolution equation numerically, we examine the critical layer thickness required for plug formation and the time taken for plugs to form, quantifying how both can be increased by increasing the capillary Bingham number.

The rest of the paper will be organised as follows. In §\ref{section:model}, we formulate the models, presenting the long-wave evolution equation in §\ref{section:LWderivation}, and the thin-film equation in §\ref{section:TFderivation}. A brief discussion of the methods for solving these equations is given in §\ref{section:solutionmethods}. Results for thin layers are presented in §\ref{section:thinfilms}. We discuss a representative numerical solution of the thin-film evolution equation in §\ref{section:TFnumsim}, and we examine the behaviour of the layer at long times in §\ref{section:longtime}. We compute and analyse static solutions of the thin-film evolution equation in §\ref{section:TFstatics}, and investigate the dependence of the evolution on the initial conditions and the capillary Bingham number in §\ref{section:dependenceTF}. Results for layers with finite thickness are presented in §\ref{section:thickfilms}. We discuss an example numerical solution of the long-wave equation in §\ref{section:LWtimeevol} and examine the dependence of the evolution on the capillary Bingham number, layer thickness and initial conditions in §\ref{section:LWdependence}, including discussion of the critical layer thickness for plug formation and the time taken to form a plug. A summary of the results, and a discussion of their significance for modelling lung airways is given in §\ref{section:discussion}.

\section{Model Formulation}\label{section:model}

\subsection{The Stokes system}\label{section:governingeqns}

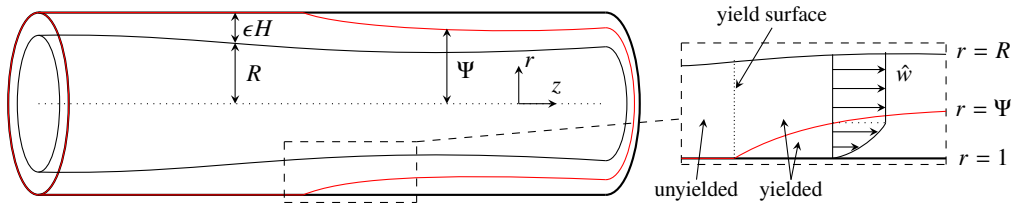
\begin{figure}
    \centering
    \begin{tikzpicture}
    \draw[thick,name path = A] (-6.5,0) ellipse (.4cm and 1.2cm);
    \draw[red,name path = C] (-6.5,0) ellipse (.4cm and 1.2cm);
    \draw[name path = B] (-6.5,0) ellipse (.28cm and .9cm);
    \draw[thick] (-6.5,1.2) -- (1,1.2);
    \draw[thick] (-6.5, -1.2) -- (1,-1.2);
    \draw[red] (-6.5, -1.2) -- (-3,-1.2);
    \draw[red] (-6.5, 1.2) -- (-3,1.2);
    \draw (-6.5,.9) .. controls (-4,.95) and (-3,0.5) .. (1,.75);
    \draw (-6.5,-.9) .. controls (-4,-.95) and (-3,-0.5) .. (1,-.75);
    \draw[red] (-3,1.2) .. controls (-2.5,1.0) and (-0.5,0.9) .. (1,1);
    \draw[red] (-3,-1.2) .. controls (-2.5,-1.0) and (-1.5,-0.9) .. (1,-1);
    \draw (1,-.75) arc(270:450:.28cm and .75cm);
    \draw[red] (1,-1) arc(270:450:.38cm and 1cm);
    \draw[thick] (1,-1.2) arc(270:450:.45cm and 1.2cm);
    \draw [stealth-stealth](-3.9,1.2) -- (-3.9,0.8);
    \draw [stealth-stealth](-3.9,0.8) -- (-3.9,0);
    \draw[dotted] (-6.5,0) -- (1,0);
    \node at (-3.65,0.4) {$R$};
    \node at (-3.6,.99) {$\epsilon H$};
    \node at (-0.85,0.45) {$\Psi$};
    \draw [stealth-stealth](-1.1,0.99) -- (-1.1,0);
    \draw[-stealth](-.15,0) -- (0.35,0);
    \draw[-stealth](-.15,0) -- (-.15,.5);
    \node at (0,.55) {$r$};
    \node at (0.35,0.205) {$z$};
    \draw[dashed] (-3.25,-.5) -- (-1.5,-.5);
    \draw[dashed] (-3.25,-.5) -- (-3.25,-1.3);
    \draw[dashed] (-3.25,-1.3) -- (-1.5,-1.3);
    \draw[dashed] (-1.5,-1.3) -- (-1.5,-.5);
    \draw[dashed] (-1.5,-.5) -- (2,-.2);
    \draw[dashed] (2,.8) -- (5.5,.8);
    \draw[dashed] (2,.8) -- (2,-.8);
    \draw[dashed] (2,-.8) -- (5.5,-.8);
    \draw[dashed] (5.5,-.8) -- (5.5,.8);
    \draw[thick] (2,-.72) -- (5.5,-.72);
    \draw (2,.5) .. controls (2.5,0.5) and (3.5,0.7) .. (5.5,.65);
    \draw[red] (2,-.72) -- (2.7,-.72);
    \draw[red] (2.7,-.72) .. controls (3.5,-.25) and (4.5,-0.15) .. (5.5,-.1);
    \draw (4,-.72) -- (4,.65);
    \draw (4,-.72) .. controls (4.4,-.65) and (4.7,-.3) .. (4.7,-.25);
    \draw (4.7,-.25) -- (4.7,.68);
    \draw[dotted] (4,-.25) -- (4.7,-.25);
    \draw[-stealth](4,.45) -- (4.7,.45);
    \draw[-stealth](4,.2) -- (4.7,.2);
    \draw[-stealth](4,-.05) -- (4.7,-.05);
    \draw[-stealth](4,-.37) -- (4.6,-.37);
    \draw[-stealth](4,-.57) -- (4.35,-.57);
    \node at (4.95,0.4) {$\hat{w}$};
    \node at (6,-.05) {$r=\Psi$};
    \node at (6,.7) {$r=R$};
    \node at (6,-.7) {$r=1$};
    \draw[densely dotted] (2.7,0.-.72) -- (2.7,0.6);
    \draw[-stealth,black](2.2,-1) -- (2.3,-.2);
    \node[scale=0.9,black] at (2.2,-1.15) {unyielded};
    \draw[-stealth,black](3.45,-1) -- (3.35,-.2);
    \draw[-stealth,black](3.45,-1) -- (3.55,-.5);
    \node[scale=0.9,black] at (3.45,-1.15) {yielded};
    \draw[-stealth,black](3.2,1) -- (2.72,.2);
    \node[scale=0.9,black] at (3.2,1.15) {yield surface};
\end{tikzpicture}
    \caption{Sketches of the geometry and flow structure in the long-wave model. In non-dimensionalised variables, the free surface is at $r=R$ and the layer thickness is $\epsilon H$. The surface $r=\Psi$ separates a region of shear-dominated flow near the cylinder wall and a region of plug-like flow near the free surface. 
    We define $\Psi\equiv\min(1,\psi)$ with $\psi$ defined in \eqref{psidefn} and $\Psi=1$ corresponding to regions of unyielded fluid. The thin-film model has qualititavely the same flow structure.}
    \label{fig:geometry}
\end{figure}

We consider a rigid circular cylinder of radius $a$ coated on the inside by a layer of Bingham fluid. The rest of the tube is filled with a gas which is assumed inviscid with spatially uniform pressure. The geometry is illustrated in figure \ref{fig:geometry}. We consider only the flow in the liquid layer. The flow is assumed to be axisymmetric, and is described by cylindrical coordinates $(r^*,z^*)$. The air-liquid interface is located at $r^* = R^*(z^*,t^*)=a-H^*(z^*,t^*)$. The fluid velocity in the film is $(u^*(r^*,z^*,t^*),w^*(r^*,z^*,t^*))$ where $u^*$ and $w^*$ are measured in the positive $r^*$ and $z^*$ directions, respectively. The non-zero components of the shear-rate tensor,
$\boldsymbol{\dot\gamma}^* = \nabla \boldsymbol{u^*}+{\nabla \boldsymbol{u^*}}^\mathrm{T}$, are therefore
\begin{equation}
    \dot\gamma^*_{rr} = 2 \p^*_ru^*, \quad
    \dot\gamma^*_{rz} = \p^*_rw^* + \p^*_zu^*, \quad
    \dot\gamma^*_{\theta\theta} = 2\frac{u^*}{r^*}, \quad
    \dot\gamma^*_{zz} = 2\p^*_zw^*.
    \label{gammadotdefn}
\end{equation}
The liquid is assumed to be incompressible and have no inertia, so the flow is governed by the Stokes equations,
\begin{subequations}
\begin{eqnarray}
        0 &=& \p^*_zw^* + \frac{1}{r^*}\partial_r^*(r^*u^*), \label{mass}
    \label{cauchy1}\\
    0 &=&  -\p^*_rp^* +
    \frac{1}{r^*}\partial_r^*(r^*\tau^*_{rr}) + \partial^*_z\tau^*_{rz} - \frac{\tau^*_{\theta\theta}}{r^*}, \label{cauchy2}\\
     0 &=& -\p^*_zp^* + \frac{1}{r^*}\partial_r^*(r^*\tau^*_{rz}) + \partial^*_z\tau^*_{zz},
     \end{eqnarray}
    \label{stokeseqns}
\end{subequations}
where $\boldsymbol{\tau}^*(r^*,z^*,t^*)$ is the stress tensor, and $p^*(r^*,z^*,t^*)$ is pressure measured relative to the gas pressure. The Bingham fluid constitutive relation is
\begin{equation}
\left. \begin{array}{ll}  
\displaystyle \tau_{ij}^* = \left(\eta + \frac{\tau_Y}{\dot\gamma^*}\right)\dot\gamma_{ij}^*
  \quad \mbox{if\ } \tau^* > \tau_Y,\\[8pt]
\displaystyle  \dot\gamma_{ij}^* = 0
  \quad \mbox{if\ } \tau^*\leq\tau_Y,
 \end{array}\right\}
  \label{stokesconstit}
\end{equation}
where $\eta$ is a viscosity, $\tau_Y$ is the yield stress, and $\dot\gamma^*$ and $\tau^*$ are the second invariants of shear-rate and stress, respectively. The second invariant of a tensor $\mathcal{T}_{ij}$ is defined as $\mathcal{T} = \sqrt{{\mathcal{T}_{ij}\mathcal{T}_{ij}}/{2}}$.

The boundary conditions are as follows. There is no slip and no penetration at the cylinder wall,
\begin{equation}
    u^* = w^* = 0  \quad \mbox{on\ }\quad r^* = a.
    \label{stokesnoslip}
\end{equation}
The kinematic boundary condition at the free surface is
\begin{equation}
    \p^*_tR^* + w^* \p^*_zR^* = u^* \quad \mbox{on\ }\quad r^*=R^*(z^*,t^*).
    \label{stokeskinBC}
\end{equation}
The gas phase applies no shear stress to the liquid film so
\begin{equation}
    -p^*n_i + \tau^*_{ij}n_j = \sigma\kappa^* n_i \quad \mbox{on\ }\quad r^*=R^*(z^*,t^*),
    \label{stressBC}
\end{equation}
where $n_i$ are the components of the normal to the free surface, $\sigma$ is the constant value of surface tension, and
\begin{equation}
    \kappa^* = \frac{1}{\sqrt{1+(\p^*_zR^*)^2}}\left[\frac{1}{R^*}-\frac{\p^*_{zz}R^*}{1+(\p^*_zR^*)^2}\right]
    \label{kappastokes}
\end{equation}
is the free-surface curvature. At the side boundaries, we impose symmetry boundary conditions,
\begin{equation}
    \p^*_zR^*=\tau^*_{rz}=w^*=0 \quad\mbox{at\ }\quad z=\{0,L^*\}.
    \label{stokessideBCs}
\end{equation}
Rather than solve the full Stokes problem defined above, we will derive reduced-order models using long-wave and thin-film theories, presented in §\ref{section:LWderivation} and §\ref{section:TFderivation}, respectively.

Surface tension at the air-liquid interface introduces an associated energy, proportional to the surface area of the air-liquid interface,
\begin{equation}
    E^* \equiv \sigma\int_0^{L^*}2\upi R^*\sqrt{1+\left(\p^*_zR^*\right)^2}\,\mathrm{d}z^*.\label{Estar}
\end{equation}
In Appendix \ref{app:stokesenergy}, we show that the Stokes equations and boundary conditions \eqref{stokeseqns}-\eqref{stokessideBCs} imply
\begin{equation}
    \p_t^*E^* = -\int_V\left(\eta\left(\dot\gamma^*\right)^2+\tau_Y\dot\gamma^*\right)\,\mathrm{d}V\leq0,
    \label{Estart}
\end{equation}
where $V$ is the volume of the layer, so the interfacial energy is always decreasing. In the Newtonian problem, the final shape that the layer reaches after its evolution can be found by solving for shapes which locally minimise interfacial energy \citep{everett_model_1972}. However, in the viscoplastic problem, this is not necessarily the case because we expect that the yield stress may freeze some or all of the layer before it has reached a minimal energy state. Analysing the final static shapes of viscoplastic layers will form a large part of our discussion, particularly in the thin-film case (cf. §\ref{section:TFstatics}). 

\subsection{The long-wave model}\label{section:LWderivation}

We non-dimensionalise the governing equations and boundary conditions \eqref{stokeseqns}-\eqref{stokessideBCs} by defining
\begin{equation}
\left. \begin{array}{l}
\displaystyle
( r, z) = \left(\frac{r^*}{a},\frac{z^*}{a}\right),\quad (u,\hat{w}) = \frac{\eta}{\sigma}\left(u^*,w^*\right), \quad R = \frac{R^*}{a}, \quad\boldsymbol{  \tau} = \frac{a}{\sigma}\boldsymbol{\tau}^*, \\[16pt]
\displaystyle

        \boldsymbol{{\dot\gamma}}=\frac{\eta a}{\sigma}\boldsymbol{\dot\gamma}^*,  \quad \hat{t}= \frac{\sigma}{a\eta}t^*,\quad  \hat{p}= \frac{a}{\sigma}p^*, \quad  \hat\kappa = a\kappa^*, \quad L = \frac{L^*}{a},
\end{array} \right\}
\label{nondim}
\end{equation}
where hats are used to distinguish \textcolor{black}{$\hat{w}$}, $\hat{t}$, $\hat{p}$ and $\hat{\kappa}$ from the scaled, thin-film quantities \textcolor{black}{$w$}, $t$, $p$ and $\kappa$ which we will define in §\ref{section:TFderivation}. After non-dimensionalising, we consider \eqref{stokeseqns}-\eqref{stokessideBCs} in a long-wave limit by introducing a characteristic axial lengthscale for the flow, $a/\delta$, where $\delta\ll1$. There is no assumption at this point that the liquid layer is thin. We then rescale the system using the small aspect ratio, $\delta$, defining 
\begin{equation}
    \bar{z}\equiv \delta z, \quad \bar{u}\equiv\frac{u}{\delta}, \quad \bar{t}\equiv \delta\hat{t}, \quad \bar{p}\equiv\delta \hat{p}, \quad \bar{L}\equiv \delta L,
    \label{scaledvars}
\end{equation}
with other variables remaining unstretched. The resulting scaled, dimensionless equations are given in Appendix \ref{app:derivation}.

To derive the long-wave evolution equation, we propose a similar flow structure in the long-wave model to that of the thin-film theory of \citet{BALMFORTH199965}. \textcolor{black}{Where the fluid is yielded, the flow is separated into a shear-dominated region, $\psi\leq r\leq1$, adjacent to the no-slip boundary, where the shear stress is large compared with the normal stresses, and a region, $ R\leq r < {\psi}$, adjacent to the free surface, where we say the flow is \say{plug-like} as the axial velocity, $\hat{w}$, is independent of $r$ (figure \ref{fig:geometry}). }We make separate expansions for the velocities and stresses in the shear-dominated and plug-like regions. \textcolor{black}{In the plug-like region, the shear stress is below the yield stress, but the fluid is still yielded because the normal stresses are large enough that the total stress exceeds the yield stress. We solve at leading order in $\delta$ in the shear-dominated and plug-like regions, match the solutions from the two regions together, and finally arrive at the evolution equation which we state below.} The full derivation is given in Appendix \ref{app:derivation}.

Following the approach of, e.g., \citet{camassa_ring_2012}, we write the evolution equation in terms of the unscaled variables \eqref{nondim} rather than the scaled variables \eqref{scaledvars}, so $\delta$ does not appear in the equations. However, the limit in which the theory is formally valid remains $\delta\ll1$.

From now on, we will use subscripts to denote derivatives. We determine the surface between the shear-dominated and plug-like regions to be
\begin{equation}
     {\psi}(z,\hat{t}) = \frac{ \hat{B}}{|\hat{p}_z|}\left(1+\sqrt{1+\left(\frac{|\hat{p}_z| {R}}{ \hat{B}}\right)^2}\right),\quad\mbox{where\ }\quad \hat{B}\equiv\frac{\tau_Ya}{\sigma} \label{psidefn}
\end{equation}
is a capillary Bingham number, which measures yield stress relative to capillary stress. We use the hat notation to distinguish $\hat{B}$ from the thin-film version, $B$, which we will introduce in §\ref{section:TFderivation}. The capillary pressure is proportional to the free-surface curvature, and is given by
\begin{equation}
    \hat{p} = -\hat{\kappa} = -\frac{1}{  R   \sqrt{1+R_z^2}}\left(1-\frac{RR_{zz}}{ 1+R_z^2}\right).\label{LWpressuregrad}
\end{equation}
We have retained the full expression for capillary pressure \eqref{LWpressuregrad} including all terms which are higher-order in $\delta$. Although this is not strictly consistent with the asymptotic analysis, it allows the evolution equation to describe capillary static effects accurately and is a widely used device  \citep[e.g.][]{gauglitz_extended_1988}. \color{black}The axial velocity is defined separately in the shear-dominated and plug-like regions,
\begin{equation}
    \hat{w} = \left\{
    \begin{array}{ll}
      \frac{1}{2}\hat{p}_z\left[\frac{1}{2}( {r}^2-1)- {\Psi}^2\log\left(r\right)\right] +  \hat{B}\sgn\left(\hat{p}_z\right)\left[ {\Psi}\log{\left(r\right)}+1-  r\right], & R\leq r < \Psi \\[2pt]
      \frac{1}{2}\hat{p}_z\left[\frac{1}{2}( \Psi^2-1)- {\Psi}^2\log\left(\Psi\right)\right] +  \hat{B}\sgn\left(\hat{p}_z\right)\left[ {\Psi}\log{\left(\Psi\right)}+1-  \Psi\right],         & \Psi \leq r \leq 1,
    \end{array} \right.
    \label{LWvelocity}
\end{equation}
where $\Psi(z,\hat{t})\equiv\min(1,\psi)$. The function $\Psi$ is defined so that \eqref{LWvelocity} applies to the whole layer, including regions of unyielded fluid. Where $\Psi=1$, the fluid is unyielded, so there is no motion and $\hat{w}=0$. The axial flux, $\hat{Q}$, is calculated by radially integrating $\hat{w}$.

\color{black}
The long-wave evolution equation is
\begin{equation}
    R_{\hat{t}}+\frac{1}{ {R}}\hat{Q}_z=0, \quad \mbox{where\ }\quad \hat{Q}= \frac{\hat{p}_z}{16}f_1(R,\Psi)+\frac{\hat{B}}{12}\sgn(\hat{p}_z)f_2(R,\Psi), \label{LWevoleqn}
\end{equation} 
with non-negative functions $f_1$ and $f_2$ (see figure \ref{fig:energy} below) given by
\begin{subequations}
\begin{eqnarray}
    f_1(R,\Psi)&\equiv& (1- {\Psi}^2)^2-2 {R}^2(1- {\Psi}^2+2 {\Psi}^2\log{ {\Psi}}),\\
    f_2(R,\Psi)&\equiv& 2-3 {\Psi}+ {\Psi}^3-6 {R}^2( {\Psi}\log{ {\Psi}}+1- \Psi).
    \end{eqnarray}
\label{f1f2defns}
\end{subequations}
The boundary conditions at the sides of the domain are
\begin{equation}
    R_z = \hat{Q} = 0 \quad\mbox{at\ }\quad z = \{0,L\}.
    \label{LWsideBCs2}
\end{equation}
The initial conditions which we impose when solving \eqref{LWevoleqn} are
\begin{equation}
    R(z,t=0) = \sqrt{(1-\epsilon)^2-\epsilon^2A^2/2}+\epsilon A\cos\left(\frac{\upi z}{L}\right),\label{LWICs}
\end{equation}
which corresponds to a flat layer perturbed by a single Fourier mode with wavelength $2L$. The constant $A$ is the perturbation amplitude and the constant $\epsilon$ is the ratio of average layer thickness to tube radius when $A=0$. The constant term in \eqref{LWICs} is chosen so that the total volume of the layer is independent of $A$ for a given $\epsilon$.

We derive an expression for the shear stress in $\Psi\leq r\leq1$ \eqref{tau0rzsheardom}, which when evaluated at $r=1$ gives the stress exerted on the tube wall,
\begin{equation}
    \hat{\tau}_w \equiv \frac{1}{2}\hat{p}_z(1-\Psi^2) + \hat{B}\sgn(\hat{p}_z)\Psi.\label{LWwallstress}
\end{equation}
Note that \eqref{LWwallstress} only holds in regions where the fluid is yielded (where $\Psi<1$). \textcolor{black}{In unyielded regions, the stress is not defined by the constitutive relation \eqref{stokesconstit} but we do know that the wall stress must be bounded by the yield stress, so $|\hat\tau_w| \leq \hat{B}$ where $\Psi=1$.}

From \eqref{Estar}, the dimensionless interfacial energy is 
\begin{equation}
    E \equiv 2\upi\int_0^LR\sqrt{1+R_z^2}\,\mathrm{d}z.
    \label{Edefn}
\end{equation}
In Appendix \ref{app:LWenergy}, we deduce directly from \eqref{LWpressuregrad}-\eqref{LWsideBCs2} that $E_{\hat{t}}\leq0$. Hence, the result \eqref{Estart} is preserved in the long-wave theory.

\subsection{The thin-film model}\label{section:TFderivation}

We now derive the analogous evolution equation for a thin film, no longer requiring $\delta\ll1$, but assuming $|1-R|\ll1$. The thin-film evolution equation can be derived using the approach of \citet{BALMFORTH199965}, but here we derive it by taking a thin-film limit of the long-wave system \eqref{psidefn}-\eqref{LWsideBCs2}. The thin-film approximation acts to flatten the geometry, so that terms in the long-wave mobility \eqref{f1f2defns} which arise due to the curvature of the geometry are negligible. The effect of the cylindrical geometry is then only felt through the free-surface curvature, which is linearised. 

We consider a layer with characteristic thickness $\epsilon a$, and now let $\epsilon\ll1$. We rescale time, defining $t\equiv\epsilon^3\hat{t}$, then define the dimensionless film thickness, $H(z,t)$, which satisfies
\begin{equation}
    R(z,\hat{t}) = 1-\epsilon H(z,t).\label{Hdefn}
\end{equation}
\textcolor{black}{It is convenient to define a radial coordinate, $y$, measured from the no-slip boundary, which satisfies $r=1-\epsilon y$. Then the free surface is located at $y=H$.} Substituting \eqref{Hdefn} into \eqref{LWpressuregrad} gives
$\hat{p}=-\hat{\kappa}=-1-\epsilon\left(H+H_{zz}\right)+O(\epsilon^2)$.
We define the thin-film curvature and capillary pressure as $\kappa \equiv ({\hat{\kappa}-1})/{\epsilon}$ and $ p \equiv ({1+\hat{p}})/{\epsilon}$,
then linearise in $\epsilon$, so that the pressure gradient driving the flow is
\begin{equation}
    p_z\equiv-\kappa_z \equiv -H_z-H_{zzz}.\label{TFkappagrad}
\end{equation}
We define the thin-film capillary Bingham number as
\color{black}
\begin{equation}
    B \equiv \frac{\hat{B}}{\epsilon^2} = \frac{\tau_Ya}{\sigma\epsilon^2}.
\end{equation}
\color{black} Expanding \eqref{psidefn}, we find $\psi = 1 - \epsilon\mathcal{Y}+O(\epsilon^2)$ where
\begin{equation}
    \mathcal{Y}\equiv H-\frac{B}{|p_z|}.\label{Ydefn}
\end{equation}
As before, we augment this definition so that it holds in yielded and unyielded regions: we define $Y\equiv\max(0,\mathcal{Y})$, which obeys $\Psi=1-\epsilon Y+O(\epsilon^2)$, with $Y=0$ corresponding to regions of unyielded fluid. \color{black}Where $Y>0$, the fluid is yielded, and the flow structure is qualitatively the same as in the long-wave model. The value of $Y$ then indicates the boundary between the shear-dominated and plug-like regions of flow. The axial velocity \eqref{LWvelocity} becomes $\hat{w} = \epsilon^3w + O(\epsilon^4)$, where the thin-film axial velocity is
\begin{equation}
    w = \left\{
    \begin{array}{ll}
      \frac{1}{2}p_z y(y-2Y), & 0\leq y < Y \\[2pt]
      -\frac{1}{2}p_zY^2,         & Y \leq y \leq H.
    \end{array} \right.\label{TFvelocity}
\end{equation}
\color{black}
Finally, substituting \eqref{Hdefn}-\eqref{Ydefn} into \eqref{LWevoleqn}-\eqref{f1f2defns}, and linearising in $\epsilon$, gives
\begin{equation}
    H_t + \frac{1}{6}\left[p_zY^2(Y-3H)\right]_z =0, \quad \mbox{where\ }\quad Y = \max\left(0,\mathcal{Y}\right) .\label{TFevoleqn}
\end{equation}
Equation \eqref{TFevoleqn}, with definitions \eqref{TFkappagrad} and \eqref{Ydefn}, is the thin-film evolution equation. 

In the thin-film limit, the boundary conditions \eqref{LWsideBCs2} become
\begin{equation}
    H_z = Q = 0 \quad\mbox{at\ }\quad z = \{0,L\},
    \label{TFsideBCs}
\end{equation}
where we define the thin-film flux as $Q\equiv p_zY^2(Y-3H)/6$. Note that in the Newtonian problem \citep[e.g.][]{hammond_nonlinear_1983}, enforcing zero flux at $z=\{0,L\}$ is equivalent to enforcing zero third derivative, $R_{zzz}=0$ or $H_{zzz}=0$. Here, the zero flux conditions \eqref{LWsideBCs2} and \eqref{TFsideBCs} are preferable because the third derivatives, $R_{zzz}$ and $H_{zzz}$, generally become discontinuous at $z=\{0,L\}$ during the evolution. This also occurs at any interior points where the direction of flow changes (cf. figure \ref{fig:snapshotsTF}c). 
This is an inconsistency in the theory which could be resolved by finding a solution in the inner region (likely of axial length $O(\epsilon)$ in the thin-film system or $O(\delta)$ in the long-wave system) around each of these points and matching these to the global outer solution which we compute. Following the approach of, e.g., \citet{balmforth_visco-plastic_2000}, we do not solve in these inner regions and assume that the solution which we compute captures the global dynamics of the layer sufficiently accurately.

After linearising in $\epsilon$ and combining with \eqref{Hdefn}, the initial condition \eqref{LWICs} becomes
\begin{equation}
    H(z,0) = 1 + A\cos{\left(\frac{\upi z}{L}\right)},
    \label{TFICs}
\end{equation}
which is the initial condition we will use when solving \eqref{TFevoleqn}. We define the thin-film wall shear stress, $\tau_w\equiv\hat{\tau}_w/\epsilon^2$, then linearise in $\epsilon$ to get
\begin{equation}
    \tau_w = p_zY + B\sgn(p_z) = p_zH,\label{TFwallstress}
\end{equation}
where we used \eqref{Ydefn} in the second equality. \textcolor{black}{Note that \eqref{TFwallstress} only holds in regions where the fluid is yielded ($Y>0$), but we have the bound $|\tau_w|\leq B$ in unyielded regions ($Y=0$).} The interfacial energy \eqref{Edefn}, when expanded in powers of $\epsilon$, becomes
\begin{equation}
    E = 2\upi L-\epsilon V_0+\upi\epsilon^2\int_0^L\left(H_z^2-H^2\right)\,\mathrm{d}z+O(\epsilon^3)
    \label{Ethin}
\end{equation}
where $V_0$ is the (constant) total volume of the layer. We show in Appendix \ref{app:TFenergy} that \eqref{TFkappagrad}, \eqref{TFevoleqn} and \eqref{TFsideBCs} imply $E_t\leq0$ for $\epsilon\ll1$. Hence, the thin-film approximation also preserves the result \eqref{Estart}.

\subsection{Solution methods}\label{section:solutionmethods}

When solving both the thin-film and long-wave equations, we choose the domain length to be $L=\sqrt{2}\upi$. This length corresponds to the half-wavelength of the most unstable mode in the Newtonian linear stability analysis \citep{hammond_nonlinear_1983}. The instability in the viscoplastic problem is inherently nonlinear, but this choice for $L$ allows direct comparison to previous literature on the Newtonian and viscoelastic versions of the problem  \citep[e.g.][]{gauglitz_extended_1988,halpern_effect_2010}. We found that small changes in $L$ do not qualitatively affect our results so $L=\sqrt{2}\upi$ can be considered a representative domain length. The form of perturbation in the initial conditions, \eqref{LWICs} or \eqref{TFICs}, then corresponds to the single unstable Fourier mode that exists in the domain. In the long-wave theory, $\delta$ is defined as the ratio of tube radius $a$ to a typical axial lengthscale. If that axial lenthscale is taken to be the wavelength of the initial disturbance, then $\delta=1/(2L)=1/(2\sqrt{2}\upi)$. Shorter wavelength structures also develop in the thin-film and long-wave simulations (cf. capillary waves discussed in §\ref{section:TFnumsim}), testing the validity of the long-wave theory. Pending validation by computations of the full problem \eqref{gammadotdefn}-\eqref{stokessideBCs}, we anticipate that our results provide a good approximation to the true behaviour, with additional accuracy gained from retaining the exact expression for $\hat{\kappa}$ in \eqref{LWpressuregrad}. \textcolor{black}{We do not observe any significantly different behaviour when applying periodic boundary conditions at the sides of the domain compared to the boundary conditions \eqref{LWsideBCs2} or \eqref{TFsideBCs}, validating our use of the latter in all the results presented. }

When solving the systems \eqref{psidefn}-\eqref{LWsideBCs2} or \eqref{TFkappagrad}-\eqref{TFsideBCs} numerically, we use the method proposed by \citet{jalaal_thesis_2016}. The evolution equations are regularised by redefining $Y\equiv\max(Y_{min},\mathcal{Y})$ and $\Psi\equiv\min(\Psi_{max},\psi)$, where $Y_{min},1-\Psi_{max}\ll1$. We choose $Y_{min}=1-\Psi_{max}=10^{-6}$ after confirming this is small enough that \textcolor{black}{the results are not sensitive to the precise value of $Y_{min}$. For example, for the simulation presented in figure 6 below, the absolute errors in $\max_zH(z,t=120)$ and $t_p$ (the time to form a plug) are bounded above by $800Y_{min}^2$ and $40000Y_{min}^2$, respectively, for all $10^{-6}\leq Y_{min}\leq10^{-3}$, which we find to be typical of convergence rates in simulations. }Where $Y=Y_{min}$ or $\Psi=\Psi_{max}$, there is a very weak regularisation-induced-flow, but since $Y_{min}$ is chosen small enough for this flow to be negligible, we treat these regions as unyielded, treating $Y=Y_{min}$ or $\Psi=\Psi_{max}$ as equivalent to $Y=0$ or $\Psi=1$. The regularised equations are solved using the method of lines: the spatial derivatives are approximated using second-order centred finite differences and the resulting system of ODEs is solved through time using a stiff solver in \textsc{Matlab}. We have confirmed that the number of spatial grid points used is large enough that the precise value does not affect our results. 

\section{Results: thin-film theory}\label{section:thinfilms}

\subsection{Time evolution of a thin layer}\label{section:TFnumsim}

\begin{figure}
\centering
\subfloat[]{\label{subfig:2a}\includegraphics[scale=1,width=.9\textwidth]{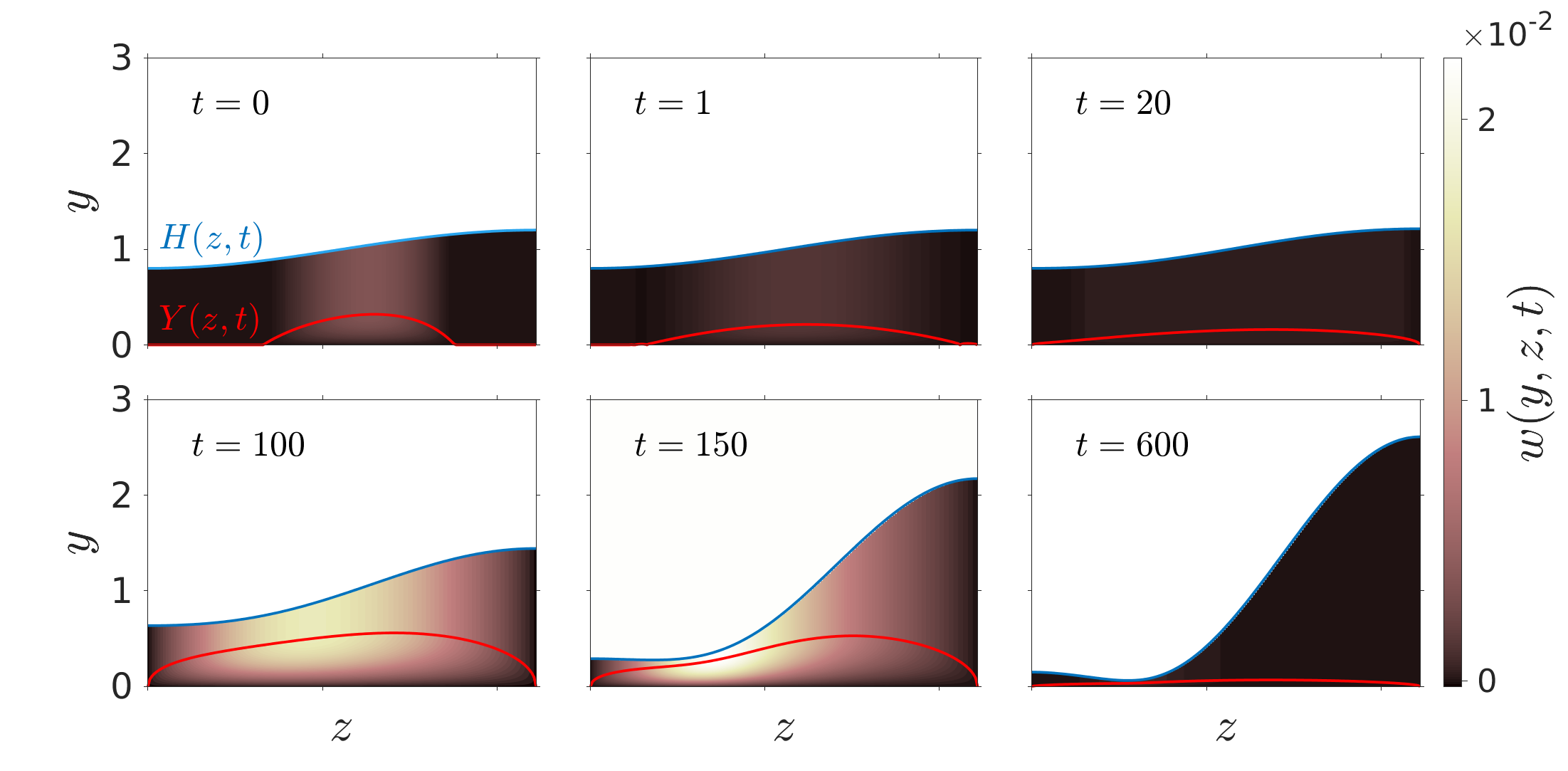}}\par\medskip
\begin{minipage}{.49\linewidth}
\centering
\subfloat[]{\label{subfig:2b}\includegraphics[scale = 0.5,width=0.9\textwidth]{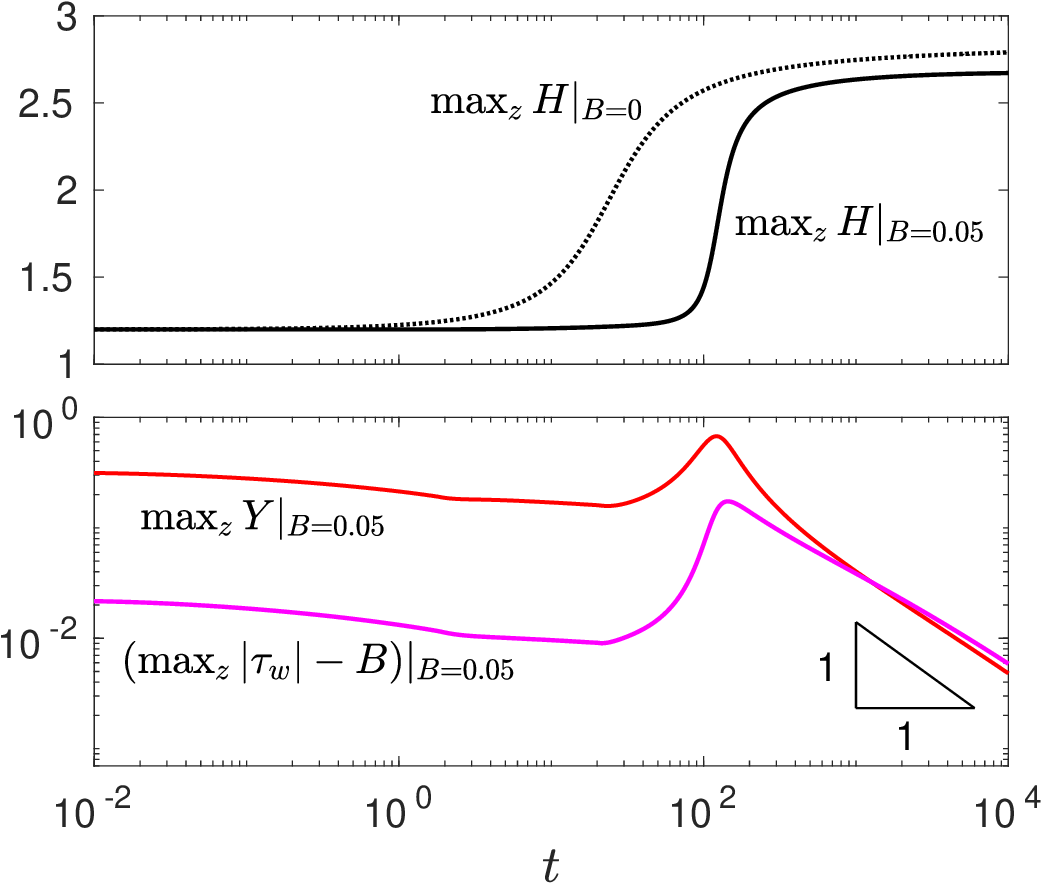}}
\end{minipage}
\begin{minipage}{.5\linewidth}
\centering
\subfloat[]{\label{subfig:2c}\includegraphics[scale=.53,width=0.98\textwidth]{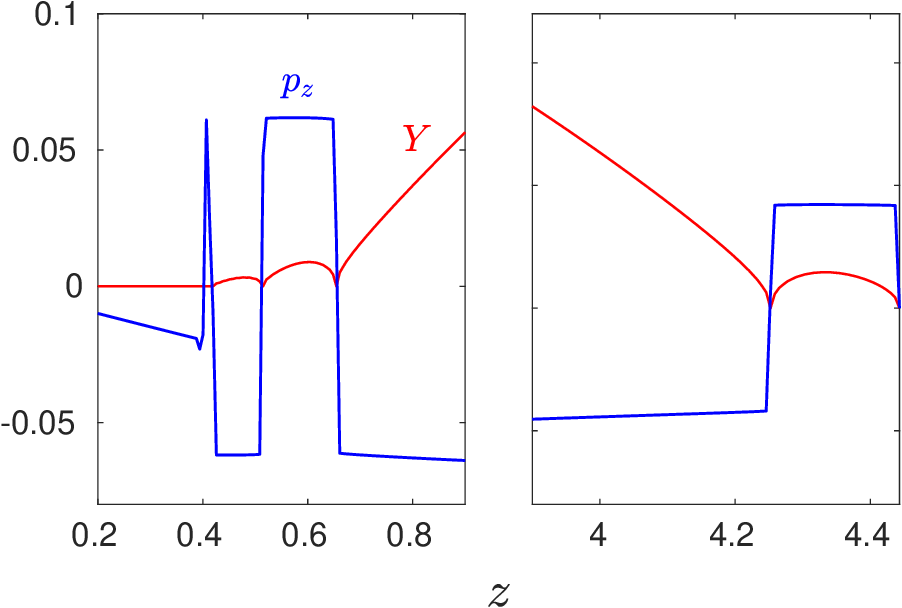}}%
\end{minipage}

\caption{Time evolution of a thin film with $B=0.05$ and $A=0.2$. (a) Snapshots of layer height $H(z,t)$ and the internal surface $Y(z,t)$ (see supplementary movie 1 for the full evolution of $H$ and $Y$ up to $t=1000$). \textcolor{black}{Axial velocity, $w$, as defined in \eqref{TFvelocity}, is also shown. The plug-like region lies between $y=Y$ and $y=H$, showing significant transient deformation.} (b) Time evolution of $\max_zH$ compared to the same quantity for a Newtonian ($B=0$) simulation, and time evolution of $\max_zY$ and $\max_z|\tau_w|-B$ for the $B=0.05$ simulation. (c) Example of the capillary-wave-like structures that are observed ahead of the travelling yield surface at early times, similar to those discussed in \citet{jalaal_stoeber_balmforth_2021}. $Y$ and $p_z$ are shown near to the travelling yield surfaces at $t=1$. The sign changes in $p_z$ indicate reversals in the direction of flow between these structures. 
}
\label{fig:snapshotsTF}
\end{figure}

Figure \ref{fig:snapshotsTF}(a) shows snapshots from a numerical solution of the thin-film equations \eqref{TFkappagrad}-\eqref{TFICs} with $B=0.05$ and $A=0.2$. For any viscoplastic simulation, the initial perturbation must be sufficiently large in order to trigger instability because a sufficiently large pressure gradient must be created to overcome the yield stress and make the fluid yield. Here, $A$ is large enough that the fluid in the centre of the domain yields, but $Y=0$ near the boundaries so the fluid there is initially rigid (figure \ref{fig:snapshotsTF}a, $t=0$). There is an initial period during which there is minimal deformation in $H$, but $Y$ deforms significantly and, by $t=20$, $Y>0$ for all $z\in(0,L)$ indicating that the whole layer is yielded. This initial yielding period causes a delay in the growth of the instability, as can be seen when $\max_zH(z,t)$ is compared to the same quantity from a Newtonian ($B=0$) simulation in figure \ref{fig:snapshotsTF}(b). After the initial yielding period, there is a period of significant deformation of the free surface. This coincides with a peak in $\max_zY(z,t)$ and a peak in the wall shear stress, $|\tau_w|$, at around $t=100$. This indicates that there is significant shear during this period, with more of the layer exhibiting shear-dominated flow and the region of plug-like flow becoming smaller.

At late times, the layer relaxes slowly towards a final marginally-yielded static shape in which $Y\rightarrow0$ across the whole layer (figure \ref{fig:snapshotsTF}(a), $t=600$). Figure \ref{fig:snapshotsTF}(b) shows that $\max_zY$ decays towards zero at a rate proportional to $t^{-1}$. From \eqref{TFwallstress}, we note that as $Y\rightarrow0$, $|\tau_w|\rightarrow B$ across the entire layer, which indicates that even when the layer reaches its final static shape, viscous and capillary effects apply a uniform stress on the tube wall equal to the yield stress. Figure \ref{fig:snapshotsTF}(b) also indicates that $\max_z|\tau_w|$ decays towards $B$ at a rate proportional to $t^{-1}$, in contrast to the Newtonian result that the peak wall shear stress decays towards zero at a rate proportional to $t^{-1/4}$ \citep{jones_film_1978,hammond_nonlinear_1983}. The late-time shape of the layer consists of a large collar of fluid around $z=L$ and a small collar around $z=0$. Figure \ref{fig:snapshotsTF}(b) shows that the final value of $\max_zH$ is lower for the $B=0.05$ solution compared to the Newtonian solution. This is because, unlike in the Newtonian evolution, not all of the fluid drains into the large collar at late times. Instead, some is trapped in the small collar so the peak height of the layer is decreased. Thus, the yield stress inhibits the growth of the instability. In §\ref{section:longtime} and §\ref{section:TFstatics}, we quantify how increasing $B$ affects the final marginally-yielded static shape of the layer. 

In the early-time period of gradual yielding, capillary-wave-like structures form ahead of the travelling yield surfaces (where $Y$ makes contact with zero). They can be identified by observing the structure of $Y$ and the pressure gradient $p_z$ (figure \ref{fig:snapshotsTF}c). There is a jump discontinuity and a change of sign in $p_z$ between each of these structures, indicating that the direction of flow reverses. At each point that $p_z$ passes through zero, we also have $Y=0$. We expect there to exist additional waves with smaller wavelengths ahead of those observed in figure \ref{fig:snapshotsTF}(c), but the numerical method can only resolve the largest few since it is limited by the size of the finite difference grid spacing. The observed structures resemble closely the capillary waves identified by \citet{balmforth_surface_2007} and \citet{jalaal_long_2016}, and studied in detail by \citet{jalaal_stoeber_balmforth_2021} in the context of spreading viscoplastic droplets. They are a feature common to surface-tension-driven viscoplastic flows, occurring when a yield surface advances into a region of unyielded fluid. Unlike in previously studied flows, here the capillary waves, in general, only exist transiently, during the early time period of gradual yielding until the whole layer yields. However, we find that for some values of $A$ and $B$ (mostly very large $A$), one or more of the jump discontinuities in $p_z$ which develop can persist for the whole evolution. In §\ref{section:dependenceTF}, we discuss how this phenomenon can affect the final static shape of the layer, and present criteria for it to occur. Until then, we focus on the case that the flow is unidirectional ($Y>0$ for all $z\in(0,L)$) after the early time capillary waves have passed. 

\subsection{Late time asymptotics for the thin-film evolution equation}\label{section:longtime}

To analyse the late time dynamics of the layer, we look for a solution in which $Y=\mathcal{Y}\rightarrow0$ as $t\rightarrow\infty$. \textcolor{black}{Numerical simulations suggest that $Y$ decays like $t^{-1}$ (figure \ref{fig:snapshotsTF}b), so we make the expansions}
\begin{equation}
    H = H_0(z;B) + \frac{H_1(z;B)}{Bt}+\dots, \quad Y = \mathcal{Y} = \frac{Y_1(z;B)}{Bt}+\dots, \quad \mbox{as\ }\quad t\rightarrow\infty.\label{longtexpansions}
\end{equation}
\textcolor{black}{The $t^{-1}$ rate of decay of $H$ towards a steady state is also consistent with the numerical results in figure 2.} For this analysis, we assume the capillary pressure is monotonic, or equivalently the pressure gradient is one-signed, $-H_{0,z}-H_{0,zzz}<0$ for all $z\in[0,L]$. This is equivalent to assuming unidirectional flow at late times. 
Substituting the expansions \eqref{longtexpansions} into the definition of $\mathcal{Y}$ \eqref{Ydefn} and the evolution equation \eqref{TFevoleqn} gives
\begin{subequations}
\begin{eqnarray}
     H_0(H_{0,z}+H_{0,zzz}) &=& B,\label{H0eqn}\\
     B(H_1-Y_1) + H_0^2(H_{1,z}+H_{1,zzz}) &=& 0, \label{H1Y1eqnA}\\
     H_1 &=& \tfrac{1}{2}\left(Y_1^2\right)_z.\label{H1Y1eqnB}
\end{eqnarray}
\end{subequations}
The boundary conditions \eqref{TFsideBCs} imply
\begin{equation}
    H_{0,z}=H_{1,z}=Y_1=0\quad\mbox{at\ }\quad z=\{0,L\},\label{longtimeBCs}
\end{equation}
and mass conservation implies
\refstepcounter{equation}
$$
    \int_0^LH_0\,\mathrm{d}z=\frac{L}{\upi}, \quad \int_0^LH_1\,\mathrm{d}z = 0.\eqno{(\theequation{\mathit{a},\mathit{b}})}\label{longtimemassBCs}
$$
From \eqref{H0eqn}, note that $H=H_0(z;B)$ is a static solution of the evolution equation \eqref{TFevoleqn} in which $\mathcal{Y}=0$ uniformly. This is not a capillary-static solution in which the pressure is everywhere uniform; instead it is a state in which the layer is uniformly marginally yielded. From \eqref{TFwallstress} we note that, in the static solution, $\tau_w=B$ uniformly, indicating that there is a stress being applied in the positive $z$-direction, but it is resisted by the yield stress, preventing flow. The functions $H_1(z;B)$ and $Y_1(z;B)$ quantify the rate at which the layer approaches the static solution at late times.

\begin{figure}
    \centering
    \includegraphics[width=\textwidth]{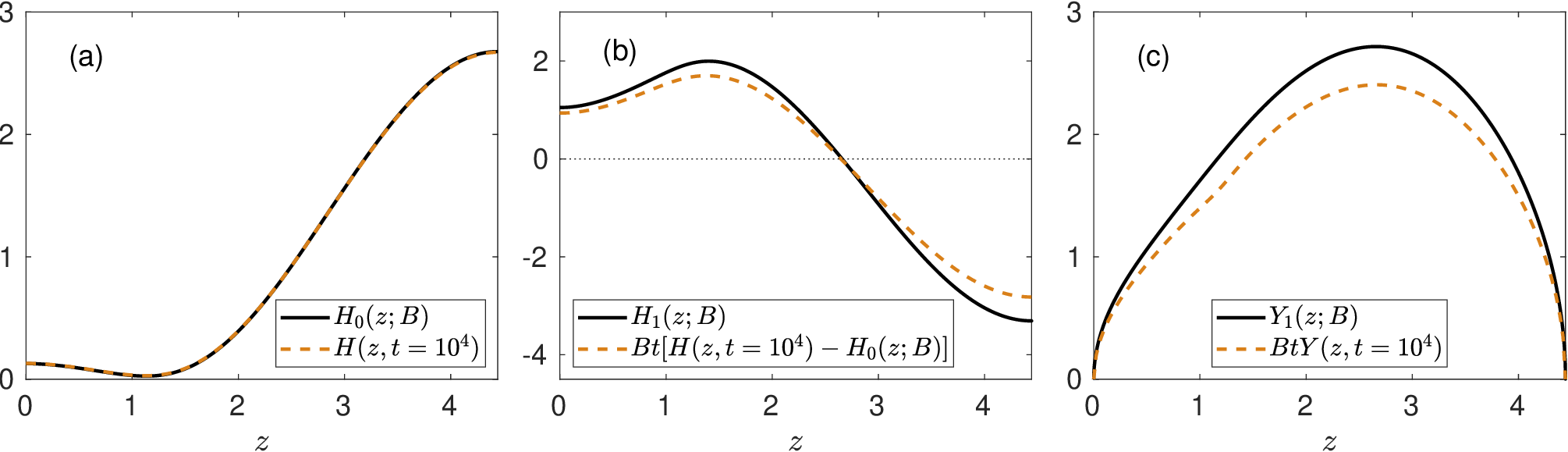}
    \caption{Late time asymptotic solutions for $B=0.05$, compared to the final snapshot of the numerical simulation with $A=0.2$ and $B=0.05$ at $t=10^4$. (a) The static solution, $H_0$, compared to the layer height, $H(z,t=10^4)$, from the simulation. (b) $H_1$ compared to $Bt[H(z,t=10^4)-H_0]$ from the simulation, which represents the rate of decay of $H$ towards $H_0$. (c) $Y_1$ compared to $BtY(z,t=10^4)$ from the simulation, which represents the rate of decay of $Y$ towards zero. 
    }\label{fig:longtimeasymp}
\end{figure}

The equations \eqref{H0eqn}-\eqref{H1Y1eqnB} were solved subject to \eqref{longtimeBCs} and \eqref{longtimemassBCs} using a boundary value problem solver in \textsc{Matlab}. A solution for $B=0.05$ is shown in figure \ref{fig:longtimeasymp} with comparison to the final snapshot of the numerical simulation from §\ref{section:TFnumsim}. Figure \ref{fig:longtimeasymp}(a) shows the agreement is very good between $H_0$ and the late-time shape of the layer from the numerical solution. Figures \ref{fig:longtimeasymp}(b) and \ref{fig:longtimeasymp}(c) show that $H_1$ and $Y_1$ approximate well the rates of decay of $H$ towards $H_0$, and $Y$ towards zero, respectively. This indicates the expansion \eqref{longtexpansions} accurately describes the late-time dynamics of the evolution and confirms the $O(t^{-1})$ rate of decay in $Y$ determined empirically in Fig \ref{fig:snapshotsTF}(b).

\subsection{Static solutions}\label{section:TFstatics}

\begin{figure}
    \centering
    \includegraphics[width=\textwidth]{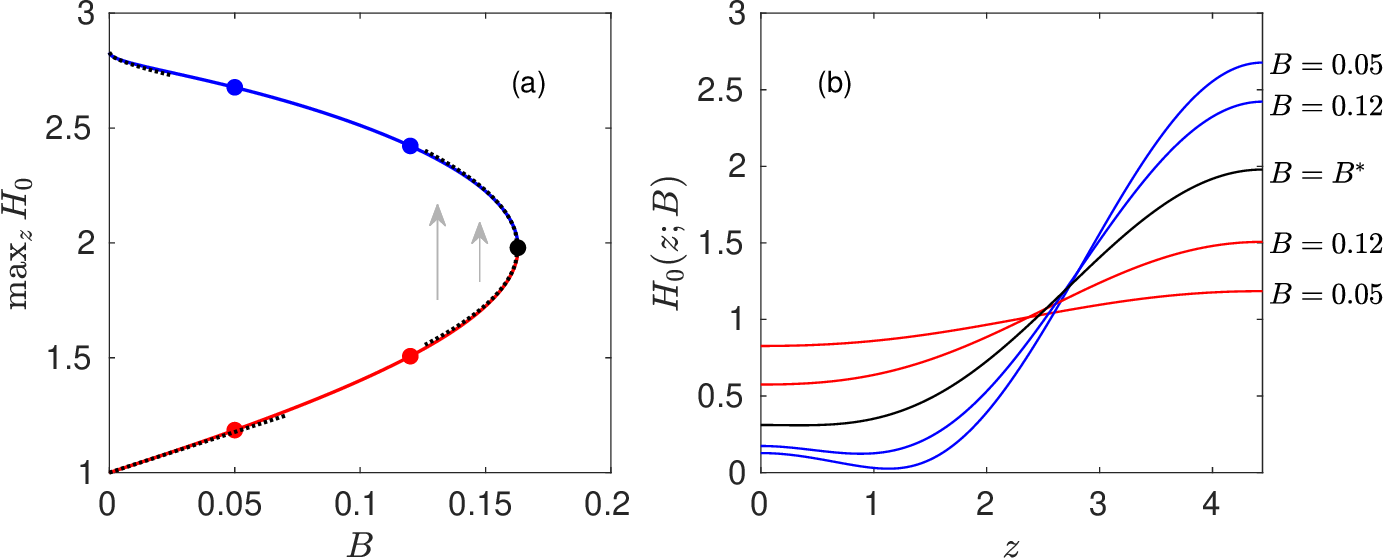}
    \caption{Solutions $H_0(z;B)$ of \eqref{H0eqn}, \eqref{longtimeBCs} and (\ref{longtimemassBCs}$\mathit{a}$), which are static solutions of the evolution equation \eqref{TFevoleqn}. (a) Bifurcation diagram showing $\max_zH_0$ for all values of $B$ such that solutions exist. Dotted black lines are the asymptotic approximations \eqref{nearbifapprox}, \eqref{lowermaxapprox2} and \eqref{uppermaxapprox}. The arrows indicate time evolution as determined by the near-bifurcation asymptotic analysis (Appendix \ref{section:nearbif}). The five dots on (a) correspond to the example solutions shown in (b), at $B=0.05$, $B=0.12$ and $B=B^*$.}
    \label{fig:bifurdiagram}
\end{figure}

The marginally-yielded static solutions, $H_0(z;B)$, can predict the final state of the layer. To investigate the dependence of $H_0$ on $B$, we solve \eqref{H0eqn} with \eqref{longtimeBCs} and (\ref{longtimemassBCs}$\mathit{a}$), varying $B$. We find that there exists a value $B^*\approx0.163$ such that for all $0\leq B< B^*$ exactly two solutions for $H_0$ exist, and for $B>B^*$ no solutions exist. There is a bifurcation at $B=B^*$. Figure \ref{fig:bifurdiagram}(a) shows $\max_zH_0(z;B)$ for all the solutions, which always coincides with $H_0(z=L;B)$. Figure \ref{fig:bifurdiagram}(b) shows several example solutions, some lying on the upper branch in figure \ref{fig:bifurdiagram}(a) and some lying on the lower branch.

The upper-branch solutions are significantly deformed layers with a large collar around $z=L$ and a small collar around $z=0$. These shapes are approached by the evolving layer at late times (figure \ref{fig:longtimeasymp}). The upper branch in figure \ref{fig:bifurdiagram}(a) quantifies the decrease in the size of the large collar formed by the layer as $B$ is increased. This decrease can be significant: $\max_zH_0$ for the upper branch ranges from the Newtonian value, $\max_zH_0(z;0)=2\sqrt{2}\approx2.83$, down to $\max_zH_0(z;B^*)\approx1.98$. This indicates that increased yield stress can significantly inhibit deformation of the film. 

The lower-branch solutions are near-flat for small $B$, becoming more deformed as $B$ is increased.
In numerical simulations, an unstable layer evolves away from a near-flat configuration towards a strongly deformed (upper-branch) static shape. This suggests that the upper-branch solutions are stable and the lower-branch solutions are unstable. We confirm the stability of the two branches using asymptotic analysis near to the bifurcation, $B\approx B^*$, presented in full in Appendix \ref{section:nearbif}. We find that $H(z,t)\sim H_0(z;B^*)+\mu\mathcal{A}(T)\phi_1(z)$ as $\mu\equiv\sqrt{B^*-B}\rightarrow0$, where $\phi_1(z)$ is a solution to a linear ODE, $T=\mu^3t$ is a slow timescale, and $\mathcal{A}(T)$ is an amplitude function which solves an ODE of the form
\begin{equation}
    \mathcal{A}_T = C_0\left(\mathcal{A}^2-\mathcal{A}_0^2\right)^2,\label{ATevolmain}
\end{equation}
where $C_0$ and $\mathcal{A}_0$ are positive constants. Equation \eqref{ATevolmain} has two fixed points, $\mathcal{A}=\pm\mathcal{A}_0$, and we compute $\mathcal{A}_0\approx2.20$. Solutions evolve away from the negative fixed point towards the positive one (figure \ref{fig:nearbif}b below), indicating that the positive fixed point is stable and the negative one unstable. These fixed points correspond to two static solutions for $H$ which are the upper- and lower-branch solutions, respectively. With this stability result, we identify the bifurcation at $B=B^*$ as a saddle-node bifurcation. We also approximate the location of the branches in figure \ref{fig:bifurdiagram}(a) by
\begin{equation}
    \max_zH_0\sim H_0(L,B^*)\pm\mu\mathcal{A}_0\phi(L)\approx 1.98\pm2.20\mu \quad \mbox{as\ } \quad \mu\equiv\sqrt{B^*-B}\rightarrow0.\label{nearbifapprox}
\end{equation} 

Since the lower-branch solutions satisfy $\mathcal{Y}=0$, they are marginal states between rigid layers ($\mathcal{Y}\leq0$) and fully yielded layers ($\mathcal{Y}>0$). We expect that if a layer is initially more deformed than the lower-branch solution it will be yielded and unstable, but if it is less deformed initially it is likely to be rigid and thus stabilised. We provide evidence from numerical simulations to confirm this in §\ref{section:dependenceTF}, where we show that the lower-branch static solutions correspond almost exactly to the minimum amplitude of initial perturbation required to trigger unstable growth.

Asymptotic analysis for small $B$ shows that the lower-branch solutions have the regular expansion,
\begin{equation}
    H_0(z;B) = 1 + B\left[x-\frac{L}{2}-\sin{x}+\frac{(1-\cos{L})\cos{x}}{\sin{L}}\right]+\dots \quad\mbox{as\ }\quad B\rightarrow0.\label{lowersmallBapproxmain}
\end{equation}
Taking the maximum value of \eqref{lowersmallBapproxmain} gives
\begin{equation}
    \max_{z}H_0= 1 + B\left[\frac{L}{2}+\cot{L}-\mathrm{cosec}\,{L}\right] +\dots\quad\mbox{as\ }\quad B\rightarrow0,
    \label{lowermaxapprox2}
\end{equation}
which approximates the lower branch in figure \ref{fig:bifurdiagram}(a). To approximate the upper-branch solutions for small $B$, we construct a solution by matched asymptotic expansions. The analysis is presented in full in Appendix \ref{section:smallB} and illustrated in figure \ref{fig:smallBMAEsoln} below. In addition to the large collar around $z=L$ and the small collar around $z=0$, we identify a third, inner region located around $z=L-\upi$ where $H_0\sim O(B^2)$. In contrast to the Newtonian problem, where the inner region is described by an ODE of the form $H^3H_{zzz}=1$ \citep{hammond_nonlinear_1983}, here the relevant ODE is of the form $HH_{zzz}=1$ (see \ref{R2eqn}). The difference arises because in the Newtonian problem there is constant flux across the inner region during the late-time draining regime, while here there must be constant stress across the inner region since $H_0$ is marginally yielded. After expanding and solving for $H_0(z,B)$ in each of the three regions and matching the solutions, a composite approximation to $H_0(z,B)$ is found. This also provides an approximation to the maximum value,
\begin{equation}
    \max_zH_0 = \frac{2L}{\upi}+4a_1B^{1/2}+\frac{4-\upi^2}{L}B+\dots\quad\mbox{as\ }\quad B\rightarrow0,\label{uppermaxapprox}
\end{equation}
where $a_1$ is a constant which depends on $L$. For $L=\sqrt{2}\upi$, we compute $a_1\approx-0.105$. Equation \eqref{uppermaxapprox} approximates the upper branch in figure \ref{fig:bifurdiagram}(a).

At the saddle-node bifurcation, $B=B^*$, the two static solutions annihilate each other, so it is only possible for the layer to select the upper-branch solution if $B<B^*$. Note that in this section, we have only computed the static solutions that have monotonic pressure, but other static solutions may exist. We show in the next section that even when $B<B^*$, the layer may select a different final static shape, depending on the initial conditions of the layer, and that it is possible to have some yielding and unstable growth for some $B>B^*$ if the initial perturbation is sufficiently large. 

\subsection{Dependence on capillary Bingham number and initial conditions}\label{section:dependenceTF}

\begin{figure}
    \centering
    \includegraphics[trim = 0 0 0 -11mm,width=\textwidth]{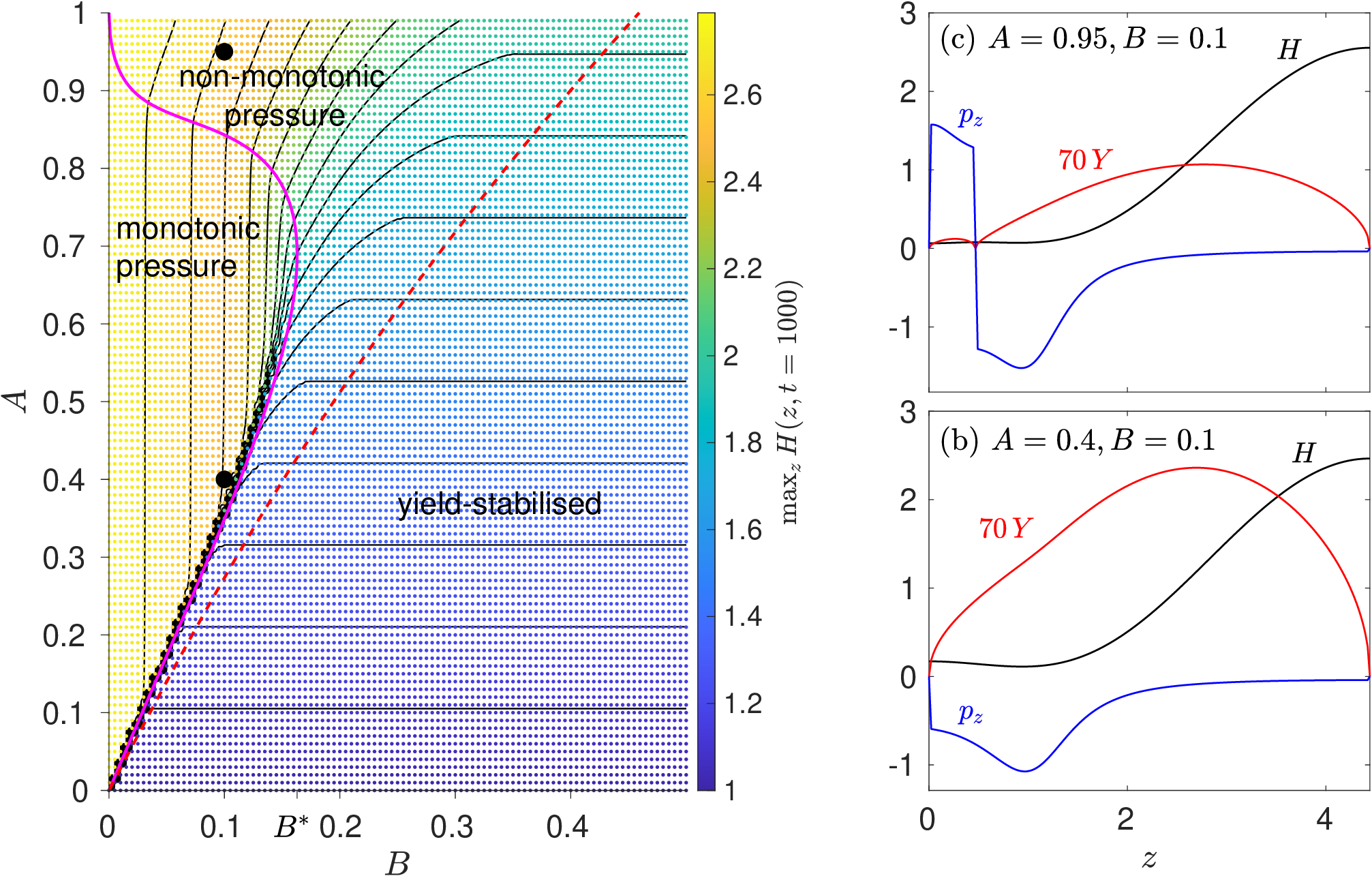}
    \caption{(a) Data from numerical solutions of the thin-film evolution equation at various $B$ and $A$. Each dot corresponds to a simulation with the colour indicating $\max H(z,t=1000)$. The data are linearly interpolated to produce the black contour lines, which are evenly spaced. When $A$ is small and $B$ is large, the layer is yield-stabilised: no unstable growth occurs. The critical amplitude for any yielding to occur, $A_m(B)$ (dashed red), defined in \eqref{Acrit}, is a strict lower bound on the boundary of the yield-stabilsed region. When there is growth, the final shape either has monotonic or non-monotonic pressure, depending on $A$ and $B$. The quantity $1-H_0(z=0,B)$ (magenta) from the static solutions (figure \ref{fig:bifurdiagram}) predicts the boundary of the monotonic pressure region. \textcolor{black}{The maximum value of $B$ along the magenta curve is $B^*$.} Two large black dots indicate the location of the example solutions shown in (b) and (c), with $H$, $70Y$ (scaled for clarity) and $p_z$ plotted at $t=1000$. (b) shows a solution in the monotonic pressure region ($p_z\leq0$) and (c) shows a solution in the non-monotonic pressure region ($p_z$ changes sign once in the domain).}
    \label{fig:AvsBthin}
\end{figure}

The final shape of the layer can depend on its initial conditions as well as $B$. To investigate this dependence, we solve the thin-film equations \eqref{TFkappagrad}-\eqref{TFICs} numerically for a range of values of $B$ and a range of initial perturbation amplitudes $A$. We run simulations on a regularly spaced grid of points in the range $(0\leq B\leq0.5,0\leq A\leq 0.99)$. 
All simulations are run to a fixed, long time, which we choose to be $t=1000$. In figure \ref{fig:AvsBthin}(a), the final maximum height, $\max_z H(z,t=1000)$, is plotted for each simulation. 

Figure \ref{fig:AvsBthin}(a) shows that there are three qualitatively different possible outcomes for an evolving layer, depending on the values of $B$ and $A$. First, there is a region for small $A$ and large $B$ with $\max_zH(z,t=1000)=1+A$, so the final maximum height is equal to the initial maximum height. In these cases, the initial perturbation does not generate a large enough pressure gradient to make the layer yield, so the yield stress entirely suppresses unstable growth. This region also extends to all of $B>0.5$ for all $0\leq A<1$. We call this the yield-stabilised region. We can seek a bound for the yield-stabilised region using the minimum amplitude, $A=A_m(B)$, such that the whole layer is initially unyielded if and only if $A<A_m$. We identify that $A_m$ is the value of $A$ such that the initial condition \eqref{TFICs} makes $\mathcal{Y}$ non-negative at exactly one point in the domain, and thus we find $A_m$ is given implicitly by
\begin{equation}
    \frac{\sqrt{1+8A_m^2}-1}{\left(4A_m^2+\sqrt{1+8A_m^2}-1\right)^{3/2}} = \frac{k|1-k^2|}{2^{5/2}B},\label{Acrit}
\end{equation}
where $k=L/\upi=\sqrt{2}/2$. The curve $A=A_m(B)$ provides a strict lower bound on the boundary of the yield-stabilised region (figure \ref{fig:AvsBthin}a). A small number of yield-stabilised simulations have $A>A_m(B)$: in these simulations the fluid initially yields a small amount in the centre of the domain but rigidifies before the sides of the domain yield, so there is no growth in $\max_zH$. 

In the second possible outcome, there is unstable growth and the layer evolves towards the upper-branch static solution with monotonic pressure which we computed in §\ref{section:TFstatics}. This occurs for a large set of $B$ and $A$ as indicated in figure \ref{fig:AvsBthin}(a), with $\max_zH(z,t=1000)$ being independent of $A$ in this region since all simulations approach the same final shape for a given $B$. Figure \ref{fig:AvsBthin}(b) shows the final shape for a simulation with $A=0.4$, $B=0.1$, which is within the monotonic pressure region. The pressure gradient $p_z$ is non-positive meaning the flow is unidirectional in the positive $z$-direction at $t=1000$ when the simulation is stopped. $Y$ is positive but close to zero for all $z\in(0,L)$ at $t=1000$, so the layer is almost rigid and $H$ is very close to the upper-branch static solution. Within the monotonic pressure region in figure \ref{fig:AvsBthin}(a), the value of $\max H(z,t=1000)$ decreases as $B$ is increased, consistent with the decrease shown in the upper branch in figure \ref{fig:bifurdiagram}(a). 

The final possible outcome involves unstable growth of the layer leading to the final static shape having non-monotonic pressure. Figure \ref{fig:AvsBthin}(a) shows that this occurs mainly when $A$ is very large. In these simulations, $\max_zH(z,t=1000)$ depends on both $B$ and $A$ and the final shape is not predicted by the upper-branch static solutions in figure \ref{fig:bifurdiagram}(a). An example of a late-time shape with non-monotonic pressure is shown in figure \ref{fig:AvsBthin}(c) from a simulation with $A=0.95$, $B=0.1$. Comparing this with figure \ref{fig:AvsBthin}(b), there is less fluid in the collar near $z=0$ and so the collar near $z=L$ is larger, even though $B$ is the same. In figure \ref{fig:AvsBthin}(c) there is exactly one point in the domain where the sign of $p_z$ changes, which corresponds to a point where the direction of flow changes. For most values of $B$ and $A$ in the non-monotonic pressure region, the final shape selected by the layer has exactly one sign-change in $p_z$ but for $A$ very close to $1$ we found that there can be more. The sign-changes in $p_z$ develop during the early-time yielding period, caused by the presence of capillary waves near the yield surfaces (figure \ref{fig:snapshotsTF}c). Most of the sign-changes exist only during this early-time period, but we see here that one or more can persist and affect the late-time dynamics, suggesting that in these cases the early-time capillary waves can influence the entire evolution. In general, the location of the sign-change(s) in $p_z$ affects the final shape of $H$, and the location of the sign-change(s) depends on the initial shape of the layer. 

There is a sharp boundary in the numerical data between the yield-stabilised and monotonic pressure regions, across which $\max_zH(z,t=1000)$ jumps significantly. There is also a clear boundary between the monotonic and non-monotonic pressure regions, indicated by where the contours of $\max_zH(z,t=1000)$ begin to curve. Figure \ref{fig:AvsBthin}(a) shows that we can predict the locations of both boundaries using the quantity $1-H_0(z=0;B)$, where $H_0(z;B)$ are the static solutions with monotonic pressure computed in §\ref{section:TFstatics}. 

First, we give an explanation for why $1-H_0(z=0;B)$ coincides with the boundary between the yield-stabilised and monotonic pressure regions. Consider a simulation with $A$ just above the boundary, so the fluid yields just enough to trigger instability (e.g. figure \ref{fig:snapshotsTF}). After the initial period of gradual yielding, $\mathcal{Y}$ is very close to zero but positive everywhere. At this point the layer's shape is very close to the lower-branch static solution with the same $B$, which has $\mathcal{Y}=0$ everywhere. Since the fluid at $z=0$ is unyielded for most of the initial period of the  evolution in which the layer gradually yields, the height at $z=0$ does not change significantly in this period. Hence, the height at $z=0$ must have initially been very close to $H_0(z=0;B)$, the height of the lower-branch static solution at that point. The initial height at $z=0$ is $H(0,0)=1-A$, so the boundary between the yield-stabilised and monotonic pressure regions is predicted by $A = 1-H_0(z=0;B)$, and the lower-branch static solutions effectively correspond to the minimum amplitude of perturbation required to trigger instability. 

Secondly, we give an explanation for why $1-H_0(z=0;B)$ coincides with the boundary between the monotonic and non-monotonic pressure regions in figure \ref{fig:AvsBthin}(a). To do this, we determine a condition for a final solution with monotonic or non-monotonic pressure to be selected during the evolution. If the initial height of the layer at $z=0$ is smaller than the height of the corresponding (i.e. for the same $B$) upper-branch static solution at $z=0$, then there must be fluid flow towards $z=0$ in the negative $z$-direction during the evolution if this solution is to be selected. This would mean that the flow at late times must not be unidirectional, and hence the pressure of the final shape must be non-monotonic. (The transient early time capillary waves in these simulations create flow reversal but it is very weak so can be neglected in this argument.) So, if $H(z=0,t=0)<H_0(z=0;B)$, for a given value of $B$, then the layer will select a final shape with non-monotonic pressure. Noting again that $H(0,0)=1-A$, we see that $A>1-H_0(z=0;B)$ is an equivalent condition for the layer to select a final shape with non-monotonic pressure.

The results in figure \ref{fig:AvsBthin}(a) are specific to the sinusoidal form of initial perturbation used \eqref{TFICs}. However, we also ran simulations with initial conditions of the form $H(z,0)=1+A\tanh{(2z-L)}$ and the results for $\max_z H(z,t=1000)$ were qualitatively, and largely quantitatively, the same. The quantity $1-H(z=0;B)$ was still found to accurately bound the monotonic pressure region. 

In this section, we have illustrated the complex dependence of the evolution of a thin layer on $B$ and $A$, and shown how the static solutions with monotonic curvature can provide insight into the dynamics of the layer. However, thin-film theory cannot capture the full range of possible dynamics for the system because the volume of fluid in a thin layer is too small to form a liquid plug in the tube. We now address this by using long-wave theory to model layers with finite thickness.

\section{Results: long-wave theory}\label{section:thickfilms}

\subsection{Time evolution of a layer with finite thickness}\label{section:LWtimeevol}

\begin{figure}
\centering
\subfloat[]{\label{main:a}\includegraphics[scale=1,width=\textwidth]{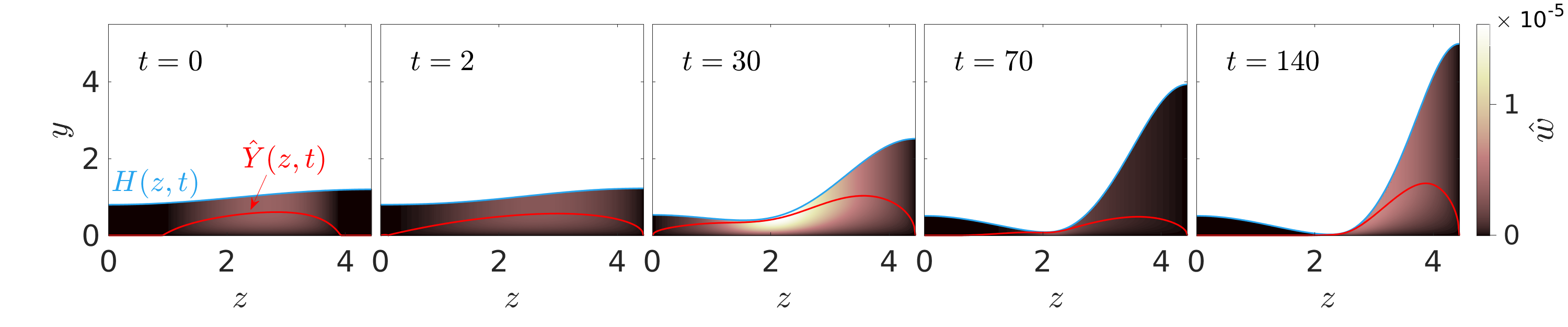}}\par\medskip
\begin{minipage}{.5\linewidth}
\centering
\subfloat[]{\label{main:b}\includegraphics[scale = 0.5,width=0.9\textwidth]{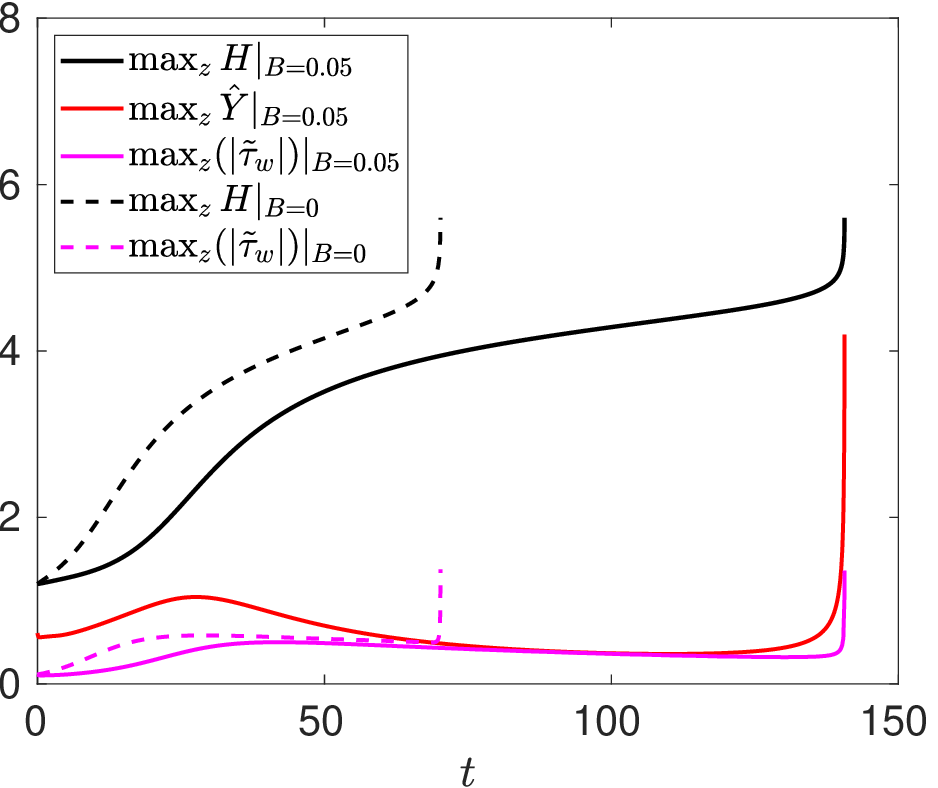}}
\end{minipage}%
\begin{minipage}{.5\linewidth}
\centering
\subfloat[]{\label{main:c}\includegraphics[scale=.5,width=0.87\textwidth]{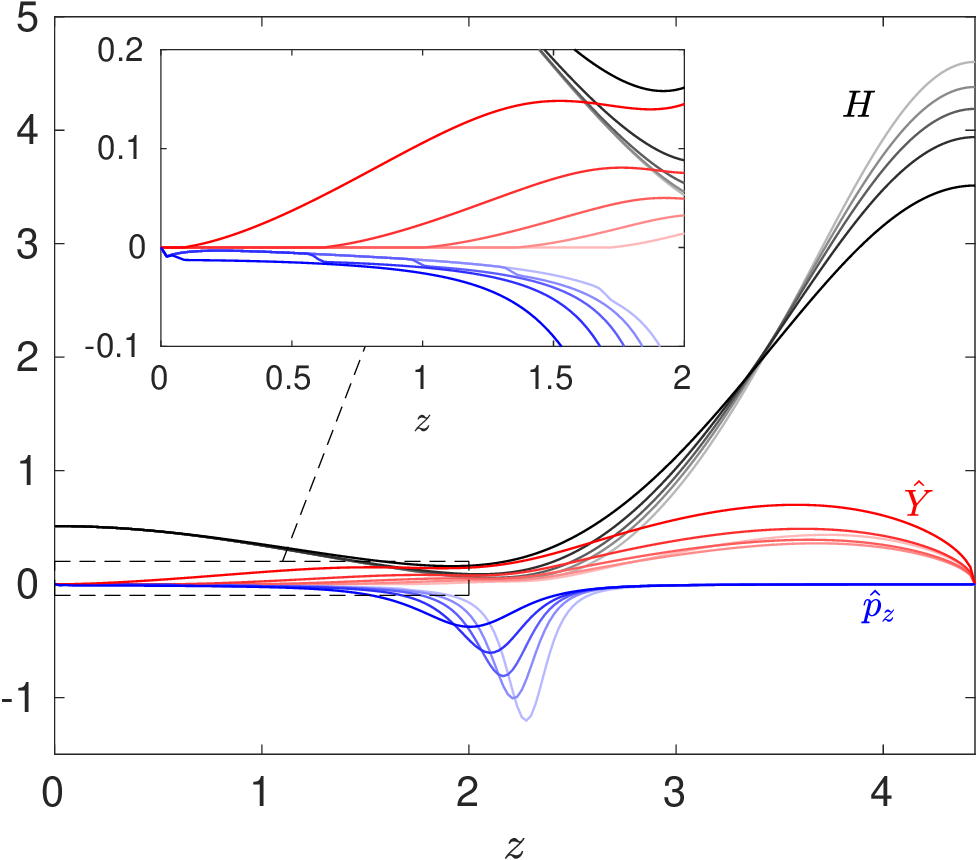}}
\end{minipage}

\caption{(a) Snapshots of a numerical simulation with $\epsilon=0.125$, $B=0.05$ and $A=0.2$ (see supplementary movie 2 for the full evolution \textcolor{black}{of $H$ and $\hat{Y}$). Axial velocity, $\hat{w}$, as defined in \eqref{LWvelocity}, is also shown.} (b) Time evolution of $\max_zH$, $\max_z\hat{Y}$ and $\max_z|\tilde{\tau}_w|$ for the same simulation, with plugging occurring at $t_p\approx140$. The simulation is stopped when $\max_zH=0.7/\epsilon$ as it is clear that a plug will form. The evolution of $\max_zH$ and $\max_z|\tilde{\tau}_w|$ for a Newtonian ($B=0$) simulation with the same $\epsilon$ and $A$ shows plug formation occurring earlier, $t_p\approx70$. (c) Snapshots of the $B=0.05$ simulation at evenly spaced time-points between $t=50$ (darkest lines) and $t=130$ (lightest lines). Inset shows $\hat{Y}$ becoming equal to zero across progressively more of the domain, indicating that the small collar of fluid around $z=0$ is rigidifying. There is a small jump in $\hat{p}_z$ where $\hat{Y}$ makes contact with zero, but still $\hat{p}\leq0$ everywhere.}
\label{fig:snapshotsLW}
\end{figure}

Figure \ref{fig:snapshotsLW}(a) shows snapshots from a numerical solution of the long-wave equations \eqref{psidefn}-\eqref{LWICs} with film thickness $\epsilon=0.125$, capillary Bingham number $B=0.05$ and initial perturbation amplitude $A=0.2$. For ease of comparison with the thin-film results, we describe solutions in terms of thin-film parameters and variables: instead of $\hat{B}$ we use $B=\hat{B}/\epsilon^2$, instead of $\hat{t}$ we use $t=\epsilon^3\hat{t}$, instead of $R(z,\hat{t})$ we use $H(z,t)$, instead of $\Psi(z,\hat{t})$ we use $\hat{Y}(z,t)\equiv(1-\Psi)/\epsilon$, and instead of $\hat{\tau}_w$ we use $\tilde{\tau}_w\equiv\hat{\tau}_w/\epsilon^2$. 

The early-time behaviour is qualitatively the same as in the thin-film simulations. There is a delay to the growth as the fluid gradually yields (figure \ref{fig:snapshotsLW}b) and capillary waves develop, which are qualitatively the same as in the thin-film case. There is then a peak in $\max_z\hat{Y}$ at around $t\approx30$, coinciding with significant deformation of the layer. Figure \ref{fig:snapshotsLW}(b) also shows that there is a small associated peak in the wall shear stress, $\max_z|{\tilde\tau}_w|$, which occurs slightly later, around $t\approx40$. The layer then evolves towards a shape with a large collar near $z=L$ and a smaller collar near $z=0$ (figure \ref{fig:snapshotsLW}a, $t=70$)

\citet{gauglitz_extended_1988} identified the critical thickness required to form a plug in their Newtonian simulations as $\epsilon=0.12$. Since $\epsilon=0.125$ is larger than this critical value, we expect plug formation may be possible in this simulation. Indeed, figure \ref{fig:snapshotsLW}(b) shows that at around $t=140$, $\max_zH$ begins to rapidly increase towards the centre of the tube, which is located at $1/\epsilon=8$. The long-wave theory cannot model coalescence so, following the approach of, e.g., \citet{halpern_effect_2010}, we stop simulations when $\max_zH=0.7/\epsilon$, but when we do run the simulation further $\max_zH$ rapidly approaches $1/\epsilon$, so it is clear that a plug will form. We denote the time taken to form a plug as $t_p$, and use the time at which we stop the simulation as a proxy for $t_p$. Figure \ref{fig:snapshotsLW}(b) shows that the plugging time, $t_p\approx140$, is significantly longer than the plugging time for a Newtonian ($B=0$) simulation, $t_p\approx70$. This is partly due to the delay caused by the initial yielding period, and partly because the rheology slows down the subsequent period of growth. Throughout the evolution, $\max_z|\tilde{\tau}_w|$ is larger for the $B=0$ simulation than for $B=0.05$, suggesting that the yield stress decreases the wall shear stress during this pre-coalescence phase of plug formation. Note that $\max_z\hat{Y}$ and $\max_z|\tilde{\tau}_w|$ increase rapidly around $t\approx t_p$, suggesting that the fluid in the large lobe is strongly yielded and the wall shear stress increases as plug formation occurs. However, we cannot expect the theory to remain accurate during this period since the assumption that radial velocities are weak no longer holds. As in the Newtonian problem \citep{johnson_nonlinear_1991}, fully two-dimensional theory is required to capture the coalescence phase of the evolution.

A new phenomenon which we have observed only in long-wave simulations is that the small collar of fluid which forms near $z=0$ can rigidify during the evolution. At $t=30$ (figure \ref{fig:snapshotsLW}a), the layer is fully yielded with $\hat{Y}>0$ for all $z\in(0,L)$. Figure \ref{fig:snapshotsLW}(c) shows that a yield surface (where $\hat{Y}$ makes contact with zero) then travels from $z=0$ at around $t=50$ to around $z=1.8$ at $t=130$, indicating that almost the entire small collar has rigidified by this point. The speed at which the yield surface travels through the domain decreases slightly through the evolution. We observe a small jump in $\hat{p}_z$ at the the point where $\hat{Y}$ becomes zero, but $\hat{p}_z$ remains non-positive everywhere including in the rigid region. We generally observe this rigidification of the small collar in our simulations whenever there is enough fluid to form a plug, i.e. $\epsilon\gtrsim0.12$. For thinner layers, $\epsilon\lesssim0.11$, we generally do not observe this phenomenon; instead, $\hat{Y}$ decays to zero from above everywhere in the domain, qualitatively the same behaviour as we observed in thin-film simulations (e.g. figure \ref{fig:snapshotsTF}). 

\subsection{Dependence on capillary Bingham number, layer thickness and initial conditions}\label{section:LWdependence}

\begin{figure}
    \centering
    \includegraphics[width=\textwidth]{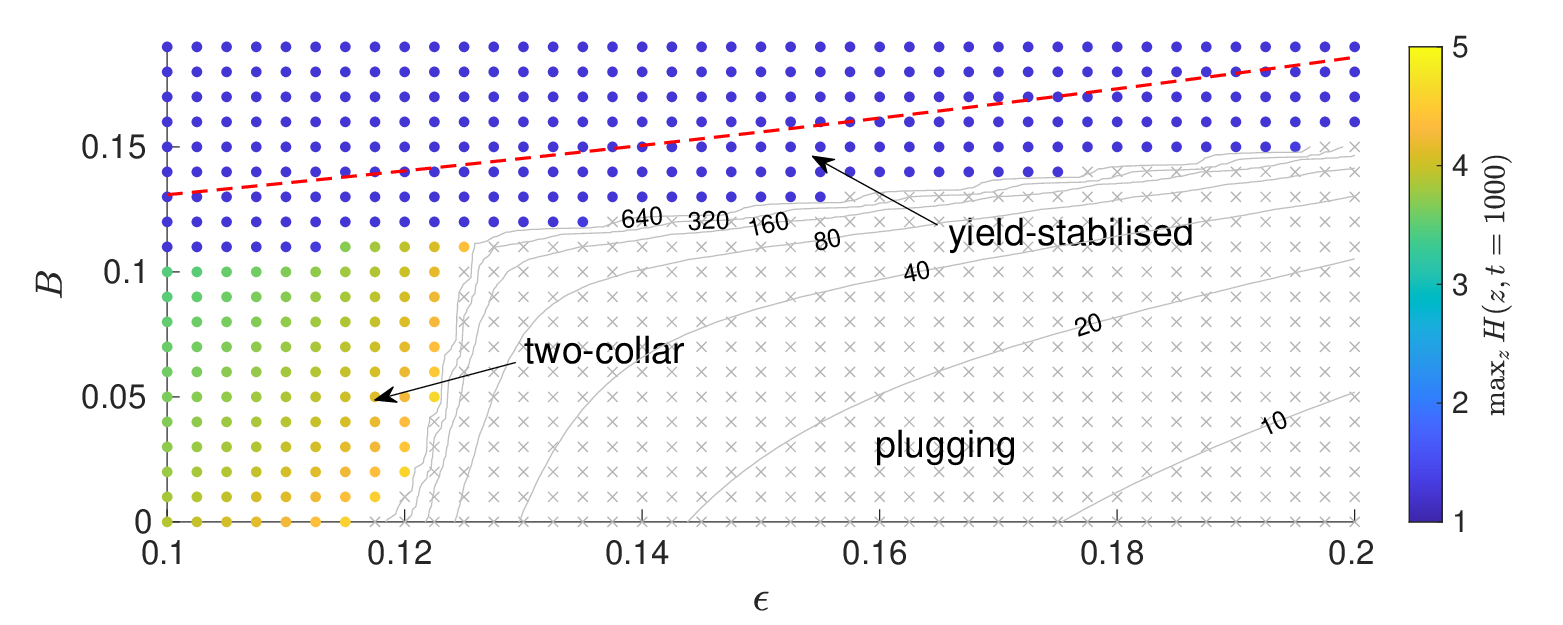}
    \caption{Data from numerical solutions of the long-wave evolution equation for various values of $B$ and $\epsilon$, with $A=0.25$. Coloured dots correspond to simulations which did not plug, with the colour indicating $\max_zH(z,t=1000)$. Grey crosses correspond to simulations which are stopped early due to a plug forming. The critical thickness $\epsilon_c$ required for plug formation can be identified as the boundary of the plugging region. The plugging time, $t_{p}$, is indicated by the grey contours. The maximum $B$ for any yielding to occur, $B_m(\epsilon,0.25)$ (red dashed), provides a strict upper bound on the yield-stabilised region.}
    \label{fig:EpsB}
\end{figure}

\begin{figure}
    \centerline{\includegraphics[width=\textwidth]{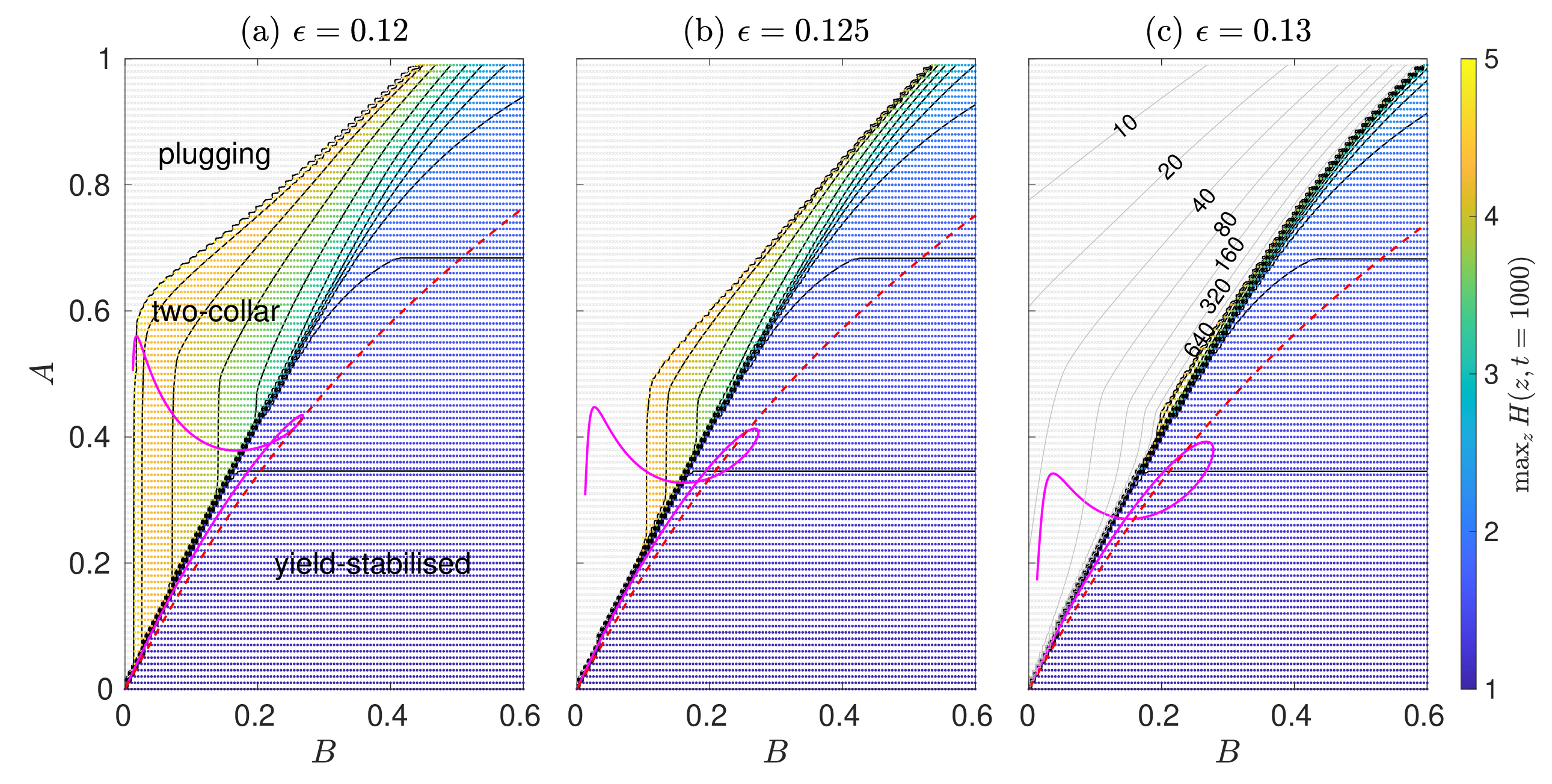}}
    \caption{Data from numerical solutions of the long-wave evolution equation for various values of $B$ and $A$, with (a) $\epsilon=0.12$, (b) $\epsilon = 0.125$, and (c) $\epsilon=0.13$. Each dot corresponds to a solution with the colour indicating $\max_zH(z,t=1000)$. The data are interpolated linearly to produce the black contour lines, which are evenly spaced. Grey points correspond to simulations which are stopped early due to a plug forming. 
    In (c), grey contours in the plugging region indicate the plugging time, $t_p$. The critical amplitude for any yielding to occur, $A_m(B,\epsilon)$ (dashed red), provides a strict lower bound on the boundary of the yield-stabilised region. Unlike in the thin-film case (figure \ref{fig:AvsBthin}a), the quantity $1-H_0(z=0;B,\epsilon)$ (solid magenta), where $H_0$ are the static solutions computed in Appendix \ref{section:LWstatics}, does not provide useful information on the final shape of these layers.}
    \label{fig:LWAvsB}
\end{figure}

We investigate the dependence of the long-wave evolution on the capillary Bingham number, layer thickness, and initial conditions by running large numbers of numerical simulations varying $B$, $\epsilon$ and $A$. This allows us to examine the dependence of the critical thickness required to form a plug, $\epsilon_c$, on $B$ and $A$. In figures \ref{fig:EpsB} and \ref{fig:LWAvsB}, we plot the final maximum heights of the layers from the numerical simulations. We run all simulations to $t=1000$, or if a plug begins to form before this time, the simulation is stopped and the stopping time is identified as $t_p$. 

In figure \ref{fig:EpsB}, we vary $\epsilon$ and $B$ between simulations while the initial perturbation amplitude is fixed at $A=0.25$. There are three distinct regions in the data, corresponding to three qualitatively different outcomes for the layer. Firstly, when $B$ is sufficiently large, there is no growth so we say the layer is yield-stabilised. As in the thin-film case in §\ref{section:dependenceTF}, we can find the minimum amplitude for yielding, $A=A_m(B,\epsilon)$, such that the layer is initially fully rigid if and only if $A\leq A_m$. This defines a corresponding capillary Bingham number, $B=B_m(\epsilon,A)$, such that for a given $A$ and $\epsilon$, the layer is fully rigid if and only if $B\geq B_m$. We compute $B_m(\epsilon,A)$ numerically by finding $B=B_m$ such that the initial condition \eqref{LWICs} makes $\psi=1$ at exactly one point in the domain, where $\psi$ is defined in \eqref{psidefn}. Figure \ref{fig:EpsB} shows that $B=B_m(\epsilon,A=0.25)$ provides a strict upper bound on the boundary of the yield-stabilised region. The second possible outcome for the layer is that there is unstable growth but a liquid plug does not form, and instead the final shape at $t=1000$ is a two-collar configuration, like in the thin-film simulations. Figure \ref{fig:EpsB} shows that this occurs for roughly $0\leq B\lesssim0.1$ and $\epsilon\lesssim0.12$, and the figure also indicates how the final peak height of the large collar, $\max_zH(z,t=1000)$, depends on both $\epsilon$ and $B$. 

In the third possible outcome, the layer forms a liquid plug, which occurs when the layer is sufficiently thick, and $B$ is not too large for it to be yield-stabilised. The boundary of the plugging region in figure \ref{fig:EpsB} corresponds to the critical thickness required for plug formation to occur, $\epsilon_c$, which can be seen to depend strongly on $B$. 
For very small $B$, the critical value is around $\epsilon_c\approx0.12$, which is consistent with the results of \citet{gauglitz_extended_1988}. As $B$ is increased, $\epsilon_c$ first increases slowly, to somewhere in the range $0.125<\epsilon_c<0.1275$ when $B=0.11$, with two-collar final shapes being observed at $\epsilon=0.125$ when $B=0.11$. This suggests that for a few values of $\epsilon$ around $\epsilon\approx0.125$, the layer can exhibit plugging, two-collar or yield-stabilised behaviour, depending on the value of $B$. For $B\geq 0.12$, the plugging region is bounded by the yield-stabilised region and $\epsilon_c$ increases rapidly as $B$ is increased. Figure \ref{fig:EpsB} also shows how the plugging time, $t_p$, decreases as $\epsilon$ is increased and increases as $B$ is increased. There is a rapid increase in $t_p$ near to the boundary of the plugging region, suggesting that we have located the boundary accurately by running simulations to $t=1000$. Simulations in the plugging region which are near the boundary spend a long time in a near-static two-collar shape (e.g. figure \ref{fig:snapshotsLW}a, $t=70$) before eventually transitioning to form a plug. 

In figure \ref{fig:LWAvsB} we investigate the dependence of the evolution on $B$ and $A$, for $\epsilon=0.12, 0.125, 0.13$. We can again identify a yield-stabilised region, a two-collar region, and a plugging region for each value of $\epsilon$. The boundary of the yield-stabilised region does not change significantly between $\epsilon=0.12$ and $\epsilon=0.13$, which is consistent with the results in figure \ref{fig:EpsB}. This boundary corresponds to the minimum amplitude of perturbation required to trigger instability, which can be seen to strongly depend on $B$. Again, we put a strict lower bound on the boundary of this region using $A_m(B,\epsilon)$, the minimum amplitude for any yielding to occur. The size of the two-collar region decreases quickly as $\epsilon$ is increased, and it has almost entirely disappeared when $\epsilon=0.13$. This suggests that for $\epsilon\geq0.13$, as long as $A$ is large enough to trigger growth, a plug is guaranteed to form. When the two-collar region does exist (figure \ref{fig:LWAvsB}a,b), the location of its boundary with the plugging region depends on $A$ and $B$. When $A$ is small, this boundary is largely independent of $A$, but when $A$ is large, the plugging region extends to higher $B$. We propose that this is because highly deformed initial conditions place most of the fluid near $z=L$, so, compared to simulations with small $A$, less fluid is trapped in the small collar near $z=0$, making the large collar larger and more unstable to plug formation. The results in figure \ref{fig:LWAvsB} show that the boundary of the plugging region depends on $A$ as well as on $B$ and $\epsilon$. Hence, the critical thickness for plug formation, $\epsilon_c$, must also depend on $A$ as well as $B$. The plugging time also depends on both $B$ and $A$, as indicated in figure \ref{fig:LWAvsB}(c), with $t_p$ increasing when $B$ is increased and decreasing when $A$ is increased. As in figure \ref{fig:EpsB}, $t_p$ increases rapidly near to the boundary of the plugging region. 

In the thin-film problem, the quantity $1-H_0(z=0;B)$ from the static solutions could be used to predict the final shape of the layer in a large number of cases (figure \ref{fig:AvsBthin}a). We have also computed static solutions, $H_0(z;B,\epsilon)$, for the long-wave problem by solving $\psi=1$ and assuming monotonic pressure (see Appendix \ref{section:LWstatics}). The quantity $1-H_0(z=0;B,\epsilon)$ is plotted for $\epsilon=0.12,0.125,0.13$ in figure \ref{fig:LWAvsB}. It predicts the threshold amplitude required to trigger instability for small $B$, but underestimates it for larger $B$, and the prediction becomes less accurate as $\epsilon$ is increased. The upper branch of the curve does not appear to be correlated with the numerical results in any way. In numerical simulations with $\epsilon\geq0.12$, we generally observe that rigidification of the small collar near $z=0$ (figure \ref{fig:snapshotsLW}b) occurs whether a plug forms or not. When this rigidification happens, the final static shape of the layer does not satisfy $\psi=1$ everywhere, so the layer selects different static solutions than the ones we have computed. The static solutions we have computed may predict the final shapes of the layer when $\epsilon$ is smaller, but they cannot do so when there is rigidification of the small collar or when there is plug formation.

The results in figures \ref{fig:EpsB} and \ref{fig:LWAvsB} are specific to the sinusoidal form of initial perturbation used \eqref{LWICs}. However, we also ran simulations with initial conditions of the form $H(z,0)=C_0+A\tanh{(2z-L)}$, where $C_0$ is a constant chosen so that the total volume of fluid is independent of $A$, and found the same qualitative behaviour and only minor quantitative differences. This suggests that the observed behaviour is not strongly dependent on the exact form of initial perturbation. 

\section{Discussion}\label{section:discussion}

To summarise, we have quantified how viscoplastic rheology can either inhibit growth of, or fully suppress, the surface-tension-driven instability of a layer of liquid coating the interior of a cylindrical tube. We found that for both thin layers and layers with finite thickness, the final shape after evolution depends sensitively on the capillary Bingham number, $B$, as well as on the initial amplitude of perturbation, $A$. Using thin-film theory, we showed that when $A$ is below a critical value, which depends on $B$, there is no unstable growth because the fluid does not yield. When there is unstable growth, the final shape of the layer either coincides with the marginally-yielded static solution, $H_0(z;B)$, from the upper branch of figure \ref{fig:bifurdiagram}(a), or the final shape has non-monotonic pressure. Figure \ref{fig:AvsBthin}(a) shows that the quantity $1-H_0(z=0;B)$ from the static solutions accurately predicts both the minimum $A$ required to trigger instability, and the large set of $A$ and $B$ for which the final shape of the layer is $H_0(z;B)$. By solving the long-wave evolution, we quantified how the critical layer thickness, $\epsilon_c$, required to form a liquid plug is increased by increasing $B$. Figure \ref{fig:EpsB} shows that $\epsilon_c$ can be increased significantly beyond the Newtonian value of $\epsilon_c\approx0.12$ found by \citet{gauglitz_extended_1988}, primarily because, when $B$ is sufficiently large, there is no yielding so no unstable growth. For $0.12\leq\epsilon\leq0.13$, it is possible for there to be no unstable growth, unstable growth leading to plug formation, or unstable growth with no plug formation, depending on the values of $A$ and $B$ (figure \ref{fig:LWAvsB}). For $\epsilon>0.13$, if $A$ is large enough to trigger instability, a plug will form. 

One application of our results is to modelling mucus flow and airway closure in lungs. We used a thin-film capillary Bingham number, $B\equiv a\tau_Y/\sigma\epsilon^2$, to measure the relative strength of the yield stress in the flow. To estimate $B$ for a $12^{\mathrm{th}}$ generation healthy airway, we propose the following typical values: airway radius $a=0.4$mm \citep{terjung_lung_2016}, surface tension $\sigma=30$mNm \citep{chen_determination_2019}, and mucus yield stress $\tau_Y=0.27$Pa \citep{patarin_rheological_2020}. For a mucus layer with thickness $\epsilon=0.125$, this gives $B\approx0.2$. Figure \ref{fig:LWAvsB}(b) shows that for $B= 0.2$ we expect plug formation to be possible, but only for $A\gtrsim0.6$, i.e. only if the mucus layer is significantly deformed initially. \citet{patarin_rheological_2020} measured the yield stress in cystic fibrosis (CF) mucus to be $6.34$Pa, which would correspond to $B\approx5$ when $\epsilon=0.125$. Figure \ref{fig:LWAvsB}(b) shows that $B=5$ is well inside the yield-stabilised region for all $A$, suggesting that airway closure would not occur via this mechanism for these parameter values. However, other key symptoms of CF are increased volume of mucus in airways and surfactant deficiency \citep{tiddens_cystic_2010}, which would correspond to increases in $\epsilon$ and $\sigma$ and, hence, a potentially significant decrease in $B$, making plug formation more likely to occur. Thus, the net effect of CF symptoms on the likelihood of airway closure by this mechanism is not obvious. Experiments and numerical modelling of plug rupture \citep{hu_microfluidic_2015,hu_effects_2020} suggest that, once an airway does close, increased yield stress makes airway reopening more difficult, which would contribute to the increased prevalence of plugged airways in CF. Our results also suggest that airway closure could be triggered if $B$ is suddenly decreased, which could be caused by applying certain therapies which are commonly used in CF, such as mucolytics which decrease the mucus yield stress \citep{patarin_rheological_2020} or expectorants which increase the volume of liquid \citep{donaldson_mucus_2006}. However, detailed modelling of the effect of such therapies would be required to confirm this conjecture. 

We have also shown that yield stress can delay plug formation when it does occur (figures \ref{fig:snapshotsLW}c, \ref{fig:EpsB} and \ref{fig:LWAvsB}c). If we set $\eta=10^{-2}\mathrm{Pa}\,\mathrm{s}$ then the dimensional plugging time is $t^*_p=(a\eta/\sigma\epsilon^3)t_p\mathrm{\, s}\approx0.07t_p\mathrm{\,s}$ for $\epsilon=0.125$. 
For the Newtonian simulation in figure \ref{fig:snapshotsLW}(c), $t_p\approx70$, corresponding to $t^*_p\approx 5\mathrm{\,s}$, which is approximately the length of a breathing cycle. The $B=0.05$ simulation in figure \ref{fig:snapshotsLW}(c) takes about twice as long to form a plug, so $t^*_p$ is likely to be longer than a breathing cycle meaning airway closure is less likely to occur via this mechanism. If we relate $\eta$ to the measured viscosity of mucus \citep{lai_micro-_2009}, $\eta=10^{-2}\mathrm{Pa}\,\mathrm{s}$ is a feasible value but it could also be significantly larger, meaning plug formation for these simulations could be on the scale of minutes or hours instead of seconds. The layer thickness also strongly influences $t_p$ (figure \ref{fig:EpsB}), and also $t^*_p$ depends inversely on $\epsilon^3$, so a modest increase in layer thickness can significantly decrease the time taken for plug formation to occur. 

Our results also suggest that the shear stress exerted on the tube wall during the pre-coalescence phase of plug formation can be decreased by yield stress (figure \ref{fig:snapshotsLW}b). This has physiological significance because a large shear stress exerted on an airway wall may cause epithelial cell damage \citep{huh_acoustically_2007}. However, we expect that the wall shear stress is likely to be much larger in the post-coalescence phase, as is the case when the liquid is Newtonian \citep{romano_liquid_2019}, so we cannot make conclusions about the effect of yield stress on wall shear stress during the entire closure process. Additionally, when the layer is too thin to form a plug, the wall shear stress at late times approaches the yield stress (figure \ref{fig:snapshotsTF}b), so in these cases it increases as yield stress is increased. 

We have focused on investigating the effect of viscoplastic rheology, so other physical effects which are relevant to airway modelling have been neglected. Various extensions to our work could be made to investigate how viscoplastic effects interact with, for example, shear stress induced by air-flow, elastic tube walls or surfactant, all of which have been studied in the case that the liquid is Newtonian \citep{halpern_fluid-elastic_1992,halpern_surfactant_1993,halpern_nonlinear_2003}. Additionally, in order to isolate the effects of the viscoplastic rheology, we have not incorporated shear-thinning or viscoelastic rheologies, which are also known to be exhibited by mucus \citep{hill_2022_mucus}. \textcolor{black}{It remains an interesting open question how the addition of other rheological properties would affect the dynamics of a viscoplastic layer as studied here. } 

There are some limitations to the thin-film and long-wave theories that we have used to derive reduced-order models. Thin-film theory cannot predict the formation of liquid plugs, and the quasi-one-dimensional long-wave theory cannot capture the fully two-dimensional dynamics which develop as a liquid plug is forming (requiring simulations to be stopped just before coalescence). Viscoplastic thin-film theory is known to break down at points where the direction of flow changes and the pressure gradient has a jump discontinuity \citep{balmforth_visco-plastic_2000}, and we observed this same behaviour in our long-wave simulations. Additionally, the long-wave theory is strictly valid for $\delta\equiv a/L\ll1$ but we solved the evolution equation in a finite domain with a small but finite value of $\delta$. We have not solved the full axisymmetric Stokes problem here, which could be used to validate the long-wave model. 

Our model predicts that viscoplastic rheology can significantly alter the evolution of a layer coating a cylindrical tube. When the layer is thin, key aspects of the dynamics and the final shape of the layer can be understood by studying marginally-yielded static solutions. When the layer has finite thickness, the critical thickness required to form a liquid plug can depend strongly on the capillary Bingham number. These results have implications for modelling real-world problems where the coating liquid has a yield stress, such as models of airway closure, particularly in the context of diseases which alter mucus rheology. 

\backsection[Supplementary data]{Supplementary movies 1 and 2 are included with the submission.}

\backsection[Acknowledgements]{The views expressed in this publication are those of the author(s) and not necessarily those of the NHS, the National Institute for Health Research, Health Education England or the Department of Health.}

\backsection[Funding]{This work was supported by the National Institute for Health Research and the NIHR Manchester Biomedical Research Centre (A.H., grant number NIHRCS12-013); the Engineering and Physical Sciences Research Council (A.B.T., grant number EP/T021365/1); and the Medical Research Council (C.A.W., grant number MR/R024944/1). J.D.S. was supported by an EPSRC Doctoral Training Award.}

\backsection[Declaration of interests]{The authors report no conflict of interest.}

\backsection[Data availability statement]{The code used to generate the data in this study is openly available in Viscoplastic Evolution Code Repository at https://doi.org/10.48420/19199756.v1.}

\backsection[Author ORCID]{J.D. Shemilt, https://orcid.org/0000-0002-9158-0930; A. Horsley, https://orcid.org/0000-0003-1828-0058; O.E. Jensen,  https://orcid.org/
0000-0003-0172-6578; A.B. Thompson,  https://orcid.org/
0000-0001-9558-1554; C.A. Whitfield,  https://orcid.org/
0000-0001-5913-735X}

\appendix

\section{Energy Evolution}\label{app:energy}

\subsection{Energy in the Stokes system}\label{app:stokesenergy}

The energy associated with the interfacial surface area is $E^*$, defined in \eqref{Estar}. Differentiating \eqref{Estar} with respect to $t^*$, integrating by parts and using the boundary condition \eqref{stokessideBCs}, gives
\begin{equation}
    \p^*_tE^* = \sigma\int_0^{L^*}2\upi{\kappa}^*R^*\p^*_tR^*\,\mathrm{d}z^*.\label{dEdt0}
\end{equation}
where $\kappa^*$ is defined in \eqref{kappastokes}. Using a standard energy balance argument for Stokes flow, such as that in \citet{Frigaard_Lectures_2019}, we can show that
\begin{equation}
    \frac{1}{2}\int_V\boldsymbol{\tau}^*\boldsymbol{:}\boldsymbol{\dot\gamma}^*\,\mathrm{d}V = \int_{\p V}\left(\boldsymbol{u}^*\boldsymbol\cdot\boldsymbol{n}\, p^*+\boldsymbol{u}^*\boldsymbol\cdot\boldsymbol{\tau}^*\boldsymbol\cdot\boldsymbol{n}\right)\,\mathrm{d}S
    \label{A2}
\end{equation}
where $V$ is the volume of the layer and $\boldsymbol{n}$ is the unit outward normal to $V$. The boundary conditions \eqref{stokesnoslip} and \eqref{stokessideBCs} imply that $\boldsymbol{u}^*\boldsymbol\cdot\boldsymbol{n}=0$ and $\boldsymbol{u}^*\boldsymbol\cdot\boldsymbol{\tau}^*\boldsymbol\cdot\boldsymbol{n}=0$ on all boundaries except $r^*=R^*$. On $r^*=R^*$, \eqref{stressBC} implies
\begin{equation}
    \boldsymbol{u}^*\boldsymbol\cdot\boldsymbol{n}\, p^*+\boldsymbol{u}^*\boldsymbol\cdot\boldsymbol{\tau}^*\boldsymbol\cdot\boldsymbol{n}=\sigma\kappa^*\boldsymbol{u}\boldsymbol\cdot\boldsymbol{n} = \sigma\kappa^*\frac{w^*\p_z^*R^*-u^*}{\sqrt{1+(\p^*_zR^*)^2}} = -\sigma\kappa^*\frac{\p_t^*R^*}{\sqrt{1+(\p_z^*R^*)^2}}\label{A3}
\end{equation}
where in the final equality we have used the kinematic boundary condition \eqref{stokeskinBC}. Substituting \eqref{A3} into \eqref{dEdt0}, then using \eqref{A2} and the fact that $\mathrm{d}S=2\upi R^*\sqrt{1+(\p_z^*R^*)^2}\mathrm{d}z^*$, we arrive at
\begin{equation}
    \p_t^*E^* = -\frac{1}{2}\int_V\boldsymbol{\tau}^*\boldsymbol{:}\boldsymbol{\dot\gamma}^*\,\mathrm{d}V.
    \label{A4}
\end{equation}
Substituting the constitutive relation \eqref{stokesconstit} into \eqref{A4} gives
\begin{equation}
    \p_t^*E^* = -\int_V\left(\eta\left(\dot\gamma^*\right)^2+\tau_Y\dot\gamma^*\right)\,\mathrm{d}V,
    \label{A5}
\end{equation}
noting that regions where $\tau^*<\tau_Y$ make no contribution to $\p_t^*E^*$ since $\boldsymbol{\dot\gamma}^*=\boldsymbol0$. Finally, since $\eta$, $\tau_Y$ and $\dot\gamma^*$ are strictly non-negative, \eqref{A5} implies $\p_t^*E^*\leq0$. 

\subsection{Energy in the long-wave system}\label{app:LWenergy}

In the rest of Appendix \ref{app:energy}, we use subscripts to denote derivatives. The non-dimensionalised expression for energy, $E$, in the long-wave system is given in \eqref{Edefn}. Differentiating \eqref{Edefn} with respect to $\hat{t}$, then inserting the evolution equation \eqref{LWevoleqn}, integrating by parts, and using the boundary conditions \eqref{LWsideBCs2}, gives
\begin{equation}
    E_{\hat{t}} = 2\upi\int_0^L\hat{p}_z\hat{Q}\,\mathrm{d}z,\label{dEdt1}
\end{equation}
where $\hat{p}$ is defined in \eqref{LWpressuregrad}. Expanding $\hat{Q}$ using the definition in \eqref{LWevoleqn} gives
\begin{equation}
    E_{\hat{t}} = -2\upi\int_0^L\left[\frac{\hat{p}_z^2}{16}f_1(R,\Psi)+\frac{\hat{B}|\hat{p}_z|}{12}f_2(R,\Psi)\right]\,\mathrm{d}z,\label{dEdt2}
\end{equation}
where the functions $f_1(R,\Psi)$ and $f_2(R,\Psi)$ are defined in \eqref{f1f2defns}. 

\begin{figure}
    \centering
    \includegraphics[width=\textwidth]{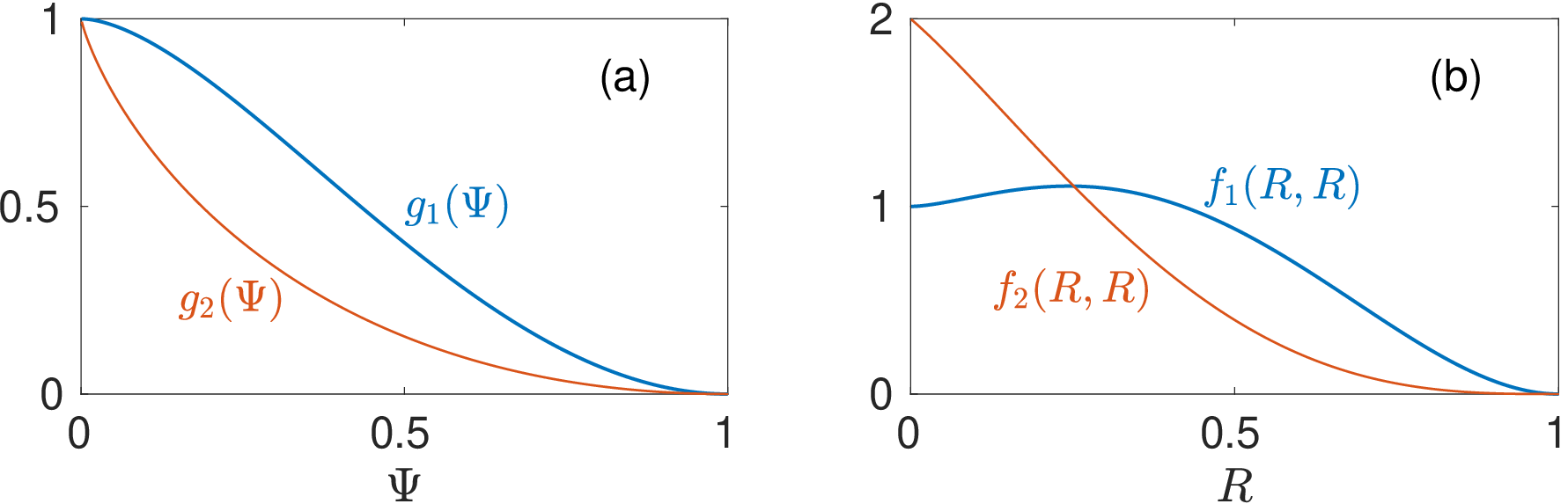}
    \caption{(a) The functions $g_1(\Psi)$ and $g_2(\Psi)$ defined in \eqref{g1g2}. (b) The functions $f_1(R,R)$ and $f_2(R,R)$ defined in \eqref{f1f2RR}. All four functions are non-negative, which is used to prove that $E_{\hat{t}}\leq0$.}
    \label{fig:energy}
\end{figure}

From \eqref{dEdt2}, $E_{\hat{t}}\leq0$ if $f_1(R,\Psi)\geq0$ and $f_2(R,\Psi)\geq0$ for all $(R,\Psi)\in \mathcal{D}\equiv\{(R,\Psi):0\leq R\leq\Psi\leq1\}$. To prove that this is the case, first note that $f_1$ and $f_2$ are both monotonically decreasing in $R$, for $(R,\Psi)\in\mathcal{D}$. This can be seen from noting that
\begin{equation}
\frac{\partial f_1}{\partial R} = -4Rg_1(\Psi),\quad
\frac{\partial f_2}{\partial R} = -12Rg_2(\Psi),\label{dfdR}
\end{equation}
where
\begin{equation}
    g_1(\Psi)\equiv1-\Psi^2+2\Psi^2\log\Psi, \quad g_2(\Psi) \equiv 1-\Psi+\Psi\log\Psi.\label{g1g2}
\end{equation}
The functions $g_1(\Psi)$ and $g_2(\Psi)$ are non-negative for all $0\leq\Psi\leq1$ (figure \ref{fig:energy}a) so the derivatives \eqref{dfdR} are both non-positive for all $(R,\Psi)\in\mathcal{D}$. This implies that if $f_1$ and $f_2$ are non-negative on the boundary $\Psi=R$ of $\mathcal{D}$, then they are non-negative everywhere in $\mathcal{D}$. Setting $\Psi=R$ in \eqref{f1f2defns}, we find the functions
\begin{subequations}
\begin{eqnarray}
    f_1(R,R) &=& (1-R^2)(1-3R^2)-4R^4\log R,\label{f1RR}\\
    f_2(R,R) &=& (R-1)(7R^2+R-2)-6R^3\log R,\label{f2RR}
\end{eqnarray}\label{f1f2RR}
\end{subequations}
are indeed non-negative for $0\leq R\leq1$ (figure \ref{fig:energy}b), so $f_1(R,\Psi)\geq0$ and $f_2(R,\Psi)\geq0$ for all $(R,\Psi)\in\mathcal{D}$. Hence, $E_{\hat{t}}\leq0$ for all admissible $(R,\Psi)$.  

\subsection{Energy in the thin-film system}\label{app:TFenergy}

We can also show directly from the thin-film equations that energy is non-increasing. The interfacial energy in the thin-film limit is \eqref{Ethin}, which when differentiated with respect to $t$ gives
\begin{equation}
    E_t \sim \upi\epsilon\int_0^L p_z^2Y^2(Y-3H)\,\mathrm{d}z \quad\mbox{as\ }\quad \epsilon\rightarrow0,
    \label{A12}
\end{equation}
where we have used integration by parts, the boundary conditions \eqref{TFsideBCs} and the evolution equation \eqref{TFevoleqn}. Noting that $0\leq Y\leq H$, \eqref{A12} immediately implies $E_t\leq0$.

\section{Derivation of the long-wave evolution equation}\label{app:derivation}

Starting from the governing equations and boundary conditions in the Stokes system \eqref{gammadotdefn}-\eqref{stokessideBCs}, we derive the long-wave evolution equation \eqref{psidefn}-\eqref{LWsideBCs2}. First, we rewrite \eqref{gammadotdefn}-\eqref{stokessideBCs} in terms of the non-dimensionalised and scaled variables \eqref{nondim} and \eqref{scaledvars}. The Stokes equations \eqref{stokeseqns} become
\begin{subequations}
\begin{eqnarray}
    0 &=&  \p_{\bar z}\hat{w} + \frac{1}{r}\partial_r(\bar ur),\label{massLW}\\
    0 &=& -\p_r\bar{p} + \delta\frac{1}{r}\partial_r(r\tau_{rr}) + \delta^2\partial_{\bar z} \tau_{rz} - \delta\tau_{\theta\theta},\label{vertmomscaled}\\
    0 &=& -\p_{\bar z}\bar{p} + \frac{1}{r}\partial_r(r\tau_{rz}) + \delta\partial_{\bar z}\tau_{zz}.\label{horizmomscaled}
\end{eqnarray}
\label{LWeqnsofmotion}
\end{subequations}
The non-zero components of the shear-rate \eqref{gammadotdefn} become
\begin{equation}
    \dot\gamma_{rr} = 2 \delta\p_r\bar u, \quad \dot\gamma_{rz} = \p_r\hat{w} + \delta^2\p_{\bar z}\bar u, \quad \dot\gamma_{\theta\theta} = 2\delta\frac{\bar u}{r}, \quad \dot\gamma_{zz} = 2\delta \p_{\bar z}\hat{w},\label{shearratecomps}
\end{equation}
and the second invariants of stress and shear-rate are
\begin{subequations}
\begin{eqnarray}
    \tau &=&  \sqrt{\frac{1}{2}(\tau_{rr}^2+\tau_{\theta\theta}^2+\tau_{zz}^2)+\tau_{rz}^2},\\
    \dot\gamma &=& \sqrt{2\delta^2\left[(\p_r\bar u)^2+\left(\frac{\bar u}{r}\right)^2+(\p_{\bar z}\hat{w})^2\right]+\left(\p_r\hat{w} + \delta^2\p_{\bar z}\bar u\right)^2}.\label{shearrate2ndinv}
\end{eqnarray}
\end{subequations}
The constitutive relation \eqref{stokesconstit} becomes
\begin{equation}
\left. \begin{array}{ll}  
\tau_{ij} = \left(1 + \frac{\hat{B}}{\dot\gamma}\right)\dot\gamma_{ij}
  \quad \mbox{if\ } \tau \geq \hat{B},\\[8pt]
\displaystyle  \dot\gamma = 0
  \quad \mbox{if\ } \tau < \hat{B},
 \end{array}\right\}
  \label{constitnondim1}
\end{equation}
where $\hat{B} \equiv {\tau_Ya}/{\sigma}$. From \eqref{stokesnoslip}, the wall boundary conditions are
\begin{equation}
    \bar u = \hat{w} = 0 \quad \mbox{on\ }\quad r = 1,
    \label{noslip}
\end{equation}
and from \eqref{stokeskinBC}-\eqref{kappastokes}, the free-surface boundary conditions are
\begin{subequations}
\begin{eqnarray}
    \p_{\bar t}R+\hat{w}\p_{\bar z}R = \bar u \quad &\mbox{on\ }&\quad r = R,
    \label{kinBC}\\
    \tau_{rz} + \delta \p_{\bar z}R(\tau_{rr}-\tau_{zz})-\delta^2(\p_{\bar z}R)^2\tau_{rz} = 0 \quad &\mbox{on\ }&\quad r = R,
    \label{shearBC}\\
    \delta\tau_{rr}-2\delta^2\p_{\bar z}R\tau_{rz}+\delta^3(\p_{\bar z}R)^2\tau_{zz} =  \left(1+\delta^2(\p_{\bar z}R)^2\right)(\bar{p}+\delta\bar\kappa)
     \quad &\mbox{on\ }& \quad
     r = R,
     \label{normalBC}
     \end{eqnarray}
\end{subequations}
where
\begin{equation}
    \bar\kappa = \frac{1}{ \sqrt{1+\delta^2(\p_{\bar z}R)^2}}\left[\frac{1}{R}-  \frac{\delta^2\p_{\bar z\bar z}R}{{1+\delta^2(\p_{\bar z}R)^2}}\right].\label{LWkappa}
\end{equation}
The symmetry boundary conditions \eqref{stokessideBCs} become
\begin{equation}
    \p_{\bar z}R = \tau_{rz} = \hat{w} = 0 \quad\mbox{at\ }\quad \bar z = \{0,\bar L\}.\label{LWsideBCs}
\end{equation}

A description of the flow where the fluid is yielded is first derived, then regions of unyielded fluid can subsequently be identified. Until otherwise stated, we assume $\tau>\hat{B}$. Separate expansions are made for the relevant variables in the shear-dominated region, $\psi\leq r\leq1$, and in the plug-like region, $R\leq r< \psi$. Quantities in the shear-dominated region are denoted by the superscript $(\cdot)^s$, and quantities in the plug-like region by $(\cdot)^p$. In the shear-dominated region, let
\begin{equation}
\left. \begin{array}{l}
\displaystyle
{\hat{w}}^s =  {w}^s_0 + \delta  {w}^s_1 + \dots,\quad {\bar u}^s =  {\bar u}^s_0 + \delta  {\bar u}^s_1 + \dots, \quad \tau^s_{rz} = \tau^s_{0rz}+\delta\tau^s_{1rz}+\dots, \\[16pt]
\displaystyle
\tau^s_{rr}=\delta\tau^s_{1rr}+\dots,\quad
        \tau^s_{zz}=\delta\tau^s_{1zz}+\dots, \quad \tau^s_{\theta\theta}=\delta\tau^s_{1\theta\theta}+\dots.
\end{array} \right\}
\label{sdexpansion}
\end{equation}
The leading-order second invariant of stress is then $\tau_0^s=|{\tau}_{0rz}^s|$. After truncating at leading order, the horizontal momentum equation \eqref{horizmomscaled} becomes
\begin{equation}
    0=-\p_{\bar z}\bar{p}+\frac{1}{r}\partial_r\left(r\tau^s_{0rz}\right)\quad \mbox{in\ }\quad\psi\leq r\leq1,
    \label{sheardomeqns}
\end{equation}
and the vertical momentum equation \eqref{vertmomscaled} implies $\bar {p}=\bar {p}(\bar z,\bar t)$ in $\psi\leq r\leq1$. The shear-dominated solution is only valid where $|\tau^s_0|>\hat{B}$, so we identify that $|\tau_0^s|=|\tau_{0rz}^s|=\hat{B}$ at $r=\psi$. After integrating \eqref{sheardomeqns} in $r$, we enforce $|\tau_{0rz}^s|=\hat{B}$ at $r=\psi$ to get
\begin{equation}
    \tau_{0rz}^s = \frac{1}{2} \p_{\bar z}\bar {p}\left(r-\frac{\psi^2}{  r}\right)+\frac{\hat{B}}{r}\sgn\left(\p_{\bar z}\bar {p}\right) \psi \quad \mbox{in\ }\quad\psi\leq r\leq1.\label{tau0rzsheardom}
\end{equation}
The shear component of the constitutive relation \eqref{constitnondim1}, at leading-order, gives
\begin{equation}
    \tau_{0rz}^s = \p_rw^s_{0} + \hat{B}\sgn\left(\p_rw^s_{0}\right) =  \p_rw^s_{0}+\hat{B}\sgn\left( \p_{\bar z}\bar {p}\right)\quad \mbox{in\ }\quad\psi\leq r\leq1.\label{tau0rzsheardom2}
\end{equation}
In both \eqref{tau0rzsheardom} and the second equality of \eqref{tau0rzsheardom2}, we have used the fact that the shear stress, $\tau^s_{0rz}$, must have the same sign as the pressure gradient. Combining \eqref{tau0rzsheardom} and \eqref{tau0rzsheardom2} gives
\begin{equation}
    \p_rw^s_{0} = \frac{1}{2}\p_{\bar z}\bar {p}\left(r-\frac{ \psi^2}{  r}\right)+  \hat{B}\sgn\left(\p_{\bar z}\bar {p}\right)\left(\frac{\psi}{r}-1\right)\quad \mbox{in\ }\quad\psi\leq r\leq1.\label{247}
\end{equation}
Integrating \eqref{247}, and enforcing the no-slip condition at $r=1$, finally gives
\begin{equation}
     {w}_0^s = \frac{1}{2}\p_{\bar z}\bar {p}\left[\frac{1}{2}( {r}^2-1)- {\psi}^2\log\left(r\right)\right] +  \hat{B}\sgn\left(\p_{\bar z}\bar {p}\right)\left[ {\psi}\log{\left(r\right)}+1-  r\right]\label{sheardomvelocity}
\end{equation}
in $\psi\leq r\leq1$.

In the plug-like region, $R\leq r<\psi$, we make an expansion of the same form as \eqref{sdexpansion},
\begin{equation}
\left. \begin{array}{l}
\displaystyle
{\hat{w}}^p =  {w}^p_0 + \delta  {w}^p_1 + \dots,\quad {\bar u}^p =  {\bar u}^p_0 + \delta  {\bar u}^p_1 + \dots, \quad \tau^p_{rz}=\tau^p_{0rz}+\delta\tau^p_{1rr}+\dots, \\[16pt]
\displaystyle
\tau^p_{rr}=\delta\tau^p_{1rr}+\dots,\quad
        \tau^p_{zz}=\delta\tau^p_{1zz}+\dots, \quad \tau^p_{\theta\theta}=\delta\tau^p_{1\theta\theta}+\dots,
\end{array} \right\}
\label{ppexpansion}
\end{equation}
but we assume the leading-order axial velocity is independent of $r$, so $w_0^p=w^p_0(\bar z,\bar t)$. This means that $\dot\gamma_{rz}=O(\delta)$, and so $\dot\gamma=O(\delta)$. Thus, the shear component of the constitutive relation \eqref{constitnondim1} implies
\begin{equation}
    \tau_{0rz}^p  = \frac{B}{\dot\gamma^p_1}\p_r w_{1}^p,\label{shearstresspp1}
\end{equation}
with \eqref{shearrate2ndinv} giving the leading-order second invariant,
\begin{equation}
    \dot\gamma_1^p = \sqrt{2\left[\left(\p_ru^{p}_{0}\right)^2+\left(\frac{u^{p}_0}{r}\right)^2+\left( \p_{\bar z}w^{p}_{0}\right)^2\right]+\left( \p_rw^{p}_{1}\right)^2}.\label{dotgamma1pp}
\end{equation}

After truncating, the horizontal momentum equation \eqref{horizmomscaled} becomes
\begin{equation}
    0=-\p_{\bar z}\bar {p}+\frac{1}{r}\partial_r\left(r\tau^p_{0rz}\right)\quad \mbox{in\ }\quad R\leq r<\psi,
    \label{pseudoplugeqns}
\end{equation}
and the vertical momentum equation \eqref{vertmomscaled} implies $\bar {p}=\bar {p}(\bar z,\bar t)$ in $R\leq r<\psi$.
The stress boundary conditions \eqref{shearBC} and \eqref{normalBC} become
\begin{equation}
    \tau^p_{0rz}=0, \quad \bar {p} = - \delta\bar {\kappa} \quad \mbox{on\ }\quad r = R.
    \label{ppBCs}
\end{equation}
Integrating \eqref{pseudoplugeqns}, and enforcing the zero shear stress condition in \eqref{ppBCs}, gives
\begin{equation}
    {\tau}_{0rz}^p = \p_{\bar z}\bar {p}\left(\frac{r}{2}-\frac{ {R}^2}{2r}\right).\label{shearstresspp2}
\end{equation}
Combining \eqref{shearstresspp1} and \eqref{dotgamma1pp}, then rearranging, gives
\begin{equation}
    \p_r{w}^p_{1} = \sqrt{2}|{\tau}^p_{0rz}|\sqrt{\frac{\left[ (\p_r{u}^p_{0})^2+({u}^p_0/r)^2+(\p_{\bar z}w^p_{0})^2 \right]}{ \hat{B}^2-( {\tau}^p_{0rz})^2}}.\label{dw1dr}
\end{equation}
The expression \eqref{dw1dr} is valid up to the point at which $\hat{B}=|\tau^p_{0rz}|$, which must coincide with $r=\psi$. Hence, using \eqref{shearstresspp2}, we arrive at the definition,
\begin{equation}
     {\psi}(\bar z,\bar t) = \frac{ \hat{B}}{|\p_{\bar z}\bar {p}|}\left(1+\sqrt{1+\left(\frac{|\p_{\bar z}\bar {p}| {R}}{ \hat{B}}\right)^2}\right).\label{psidefnapp}
\end{equation}
Note that from \eqref{psidefnapp}, $R\leq\psi$ always holds. 

The axial velocity in the shear-dominated region \eqref{sheardomvelocity} is matched to $w^p_0$ by equating them at $r=\psi$, which gives
\begin{equation}
     {w}^p_0 = \frac{1}{2}\p_{\bar z}\bar {p}\left[\frac{1}{2}( {\psi}^2-1)- {\psi}^2\log( {\psi})\right]+ {\hat{B}}\sgn\left(\p_{\bar z}\bar {p}\right)\left[ {\psi}\log( {\psi})+1-{\psi}\right]\label{w0p}
\end{equation}
in $R\leq r<\psi$. Since $\bar {p}=\bar {p}(\bar z,\bar t)$, \eqref{ppBCs} implies $\bar {p} = -\delta\bar {\kappa}$. Then \eqref{sheardomvelocity} and \eqref{w0p} provide the complete expression for axial velocity, $w_0$, across the whole layer, $ {R}\leq {r}\leq1$, in regions where the fluid is yielded. Using the same argument as \citet{BALMFORTH199965}, we identify that if $\psi(\bar z,\bar t)\geq1$ then the shear-dominated region does not exist, and the boundary conditions \eqref{noslip} imply the plug-like region must be stationary, $w^p_0=w^p_1=0$, so the fluid is unyielded. If we replace $\psi$ with $\Psi(\bar z,\bar t)\equiv\min(1,\psi)$ in \eqref{sheardomvelocity} and \eqref{w0p}, then $w_0=w_0^p$ for $R\leq r<\Psi$ and $w_0=w_0^s$ for $\Psi\leq r\leq1$ hold both where the fluid is yielded and where it is unyielded. The leading-order axial flux is then
\begin{equation}
    \hat{Q} \equiv \int_R^\Psi w^p_0\,r\,\mathrm{d}r + \int_\Psi^1 w^s_0\,r\,\mathrm{d}r,
\end{equation}
which when evaluated gives the expression in \eqref{LWevoleqn}. Finally, the kinematic boundary condition \eqref{kinBC} and mass conservation \eqref{massLW} are combined to give the evolution equation \eqref{LWevoleqn}. When we present the evolution equation and relevant definitions in \eqref{psidefn}-\eqref{LWsideBCs2}, we write them in terms of the unscaled variables $z$, $\hat{t}$, $\hat{p}$, $\hat{\kappa}$, $u$, i.e. we view the system in a frame unscaled by the small aspect ratio $\delta$. However, the limit in which the theory is formally valid remains $\delta\ll1$.

\section{Dynamics near the bifurcation in the thin-film static solutions}\label{section:nearbif}

We consider the dynamics of $H(z,t)$, as governed by the thin-film system \eqref{TFkappagrad}-\eqref{TFsideBCs}, near to $B=B_*$, the location of the saddle-node bifurcation in the static solutions computed in §\ref{section:TFstatics}. We define $\mu$ such that $B = B_*-\mu^2$ and then expand
\begin{equation}
    H = \mathcal{H}_0+\mu \mathcal{H}_1 + \mu^2 \mathcal{H}_2 + \dots, \quad Y = \mathcal{Y}_0 + \mu \mathcal{Y}_1 + \mu^2 \mathcal{Y}_2 + \dots \quad\mbox{as\ }\quad \mu\rightarrow0. \label{nearbifexpansions}
\end{equation}
We assume monotonic capillary pressure, so $|p_z|=H_z+H_{zzz}$, and consider situations where the layer is fully yielded, so $Y=\mathcal{Y}=H-B/(H_z+H_{zzz})$. We insert the expansions \eqref{nearbifexpansions} into this definition of $\mathcal{Y}$, and solve at each order in $\mu$. 

\begin{figure}
    \centering
    \includegraphics[width=\textwidth]{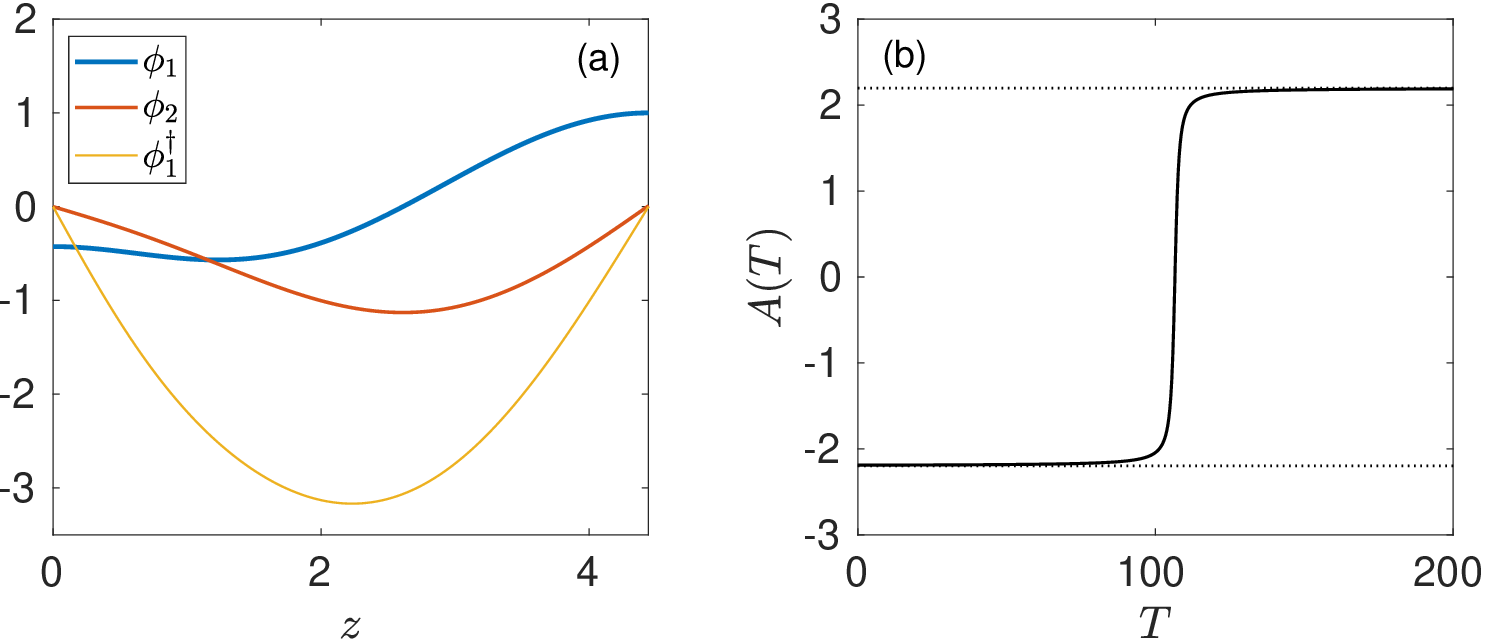}
    \caption{(a) Solutions $\phi_1$, $\phi_2$ of the linear problem $\mathcal{L}\boldsymbol\phi=\boldsymbol{0}$, and the solution $\phi_1^\dagger$ to the adjoint problem $\mathcal{L}^\dagger\boldsymbol\phi^\dagger=\boldsymbol{0}$. We set the amplitude of $\phi_1$ by choosing $\phi_1(L)=1$ and the amplitude of $\phi_1^\dagger$ by choosing $\phi_2^\dagger=1$, both of which are arbitrary values. (b) Solution of the evolution equation \eqref{ATevoleqn}, showing $\mathcal{A}$ evolving away from the fixed point at $-2.20$ towards the fixed point at $2.20$.}
    \label{fig:nearbif}
\end{figure}

Solving at $O(\mu^0)$ we get $\mathcal{Y}_0=0$ and $\mathcal{H}_0(\mathcal{H}_{0,z}+\mathcal{H}_{0,zzz})=B_*$, so $\mathcal{H}_0=H_0(z,B_*)$. The function $H_0(z,B_*)$ was plotted in figure \ref{fig:bifurdiagram}(b). To solve at $O(\mu)$, we look for a solution with $\mathcal{Y}_1=0$. This gives
\begin{equation}
    \mathcal{H}_{1,zzz}+\mathcal{H}_{1,z}+\mathcal{G}\mathcal{H}_1=0,\quad\mbox{where\ }\quad \mathcal{G}\equiv \frac{\mathcal{H}_{0,zzz}+\mathcal{H}_{0,z}}{\mathcal{H}_0}.\label{H1nearbifeqn}
\end{equation}
The associated boundary conditions are $\mathcal{H}_{1,z}=0$ at $z=\{0,L\}$, and mass conservation implies $\int_0^L\mathcal{H}_{1}\,\mathrm{d}z=0$. We look for a separable solution of \eqref{H1nearbifeqn} of the form $\mathcal{H}_1=\mathcal{A}(t)\phi_1(z)$. The function $\phi_1(z)$ is found by solving the following linear ODE problem. Defining the linear and boundary operators
\begin{equation}
    \mathcal{L} = \begin{pmatrix}\label{LBopdefn}
\partial_z^3+\partial_z+\mathcal{G} & 0 \\
1 & -\partial_z
\end{pmatrix}
\quad\mbox{and\ }\quad
    \mathcal{B} = \begin{pmatrix}
    \partial_z|_{0,L} & 0\\
    0 & \cdot|_{0,L}
    \end{pmatrix},
\end{equation}
the vector $\boldsymbol\phi=(\phi_1,\phi_2)^T$ is the solution to $\mathcal{L}\boldsymbol\phi=\boldsymbol0$ with boundary conditions $\mathcal{B}\boldsymbol\phi=\boldsymbol0$. Figure \ref{fig:nearbif}(a) shows the computed solution. Note that the amplitude of $\phi_1$ is free, so to solve we choose an arbitrary value by setting $\phi_1(L)=1$. 

We now note the linear ODE problem defined by \eqref{LBopdefn} has an associated adjoint problem. The adjoint operator, $\mathcal{L}^\dagger$, and boundary operator, $\mathcal{B}^\dagger$, are defined via the relation $\langle\boldsymbol\phi^\dagger,\mathcal{L}\boldsymbol\phi\rangle=\langle\mathcal{L}^\dagger\boldsymbol\phi^\dagger,\boldsymbol\phi\rangle$, where the inner product is defined as  $\langle\boldsymbol{\psi},\boldsymbol{\chi}\rangle=\int_0^L\boldsymbol{{\psi}}^T\boldsymbol{\chi}\,\mathrm{d}z$ for vectors $\boldsymbol\psi$, $\boldsymbol\chi$. This gives
\begin{equation}
    \mathcal{L}^\dagger = \begin{pmatrix}
-\partial_z^3-\partial_z+\mathcal{G} & 1 \\
0 & \partial_z
\end{pmatrix}
\quad\mbox{and\ }\quad
    \mathcal{B}^\dagger = \begin{pmatrix}
    \cdot|_{0,L} & 0\\
    \partial_z^2|_{0,L} & 0
    \end{pmatrix}.
    \label{adjointproblem}
\end{equation}
The adjoint solution, $\boldsymbol{\phi}^\dagger=(\phi_1^\dagger,\phi_2^\dagger)^T$, satisfies $\mathcal{L}^\dagger\boldsymbol\phi^\dagger=\boldsymbol0$ with boundary conditions $\mathcal{B}^\dagger\boldsymbol\phi^\dagger=\boldsymbol0$. The function $\phi_2^\dagger$ is constant; its value is arbitrary but sets the amplitude of $\phi_1^\dagger$. Figure \ref{fig:nearbif}(a) shows the computed solution for $\phi_1^\dagger$ where we set $\phi_2^\dagger=1$. 

To find $\mathcal{A}(t)$, first note that the evolution equation \eqref{TFevoleqn} implies $\mu \mathcal{H}_{1t}=O(\mu^4)$. We introduce a slow timescale $T=\mu^3t$ and let $\mathcal{A}=\mathcal{A}(T)$. An evolution equation for $\mathcal{A}(T)$ is found by solving the $O(\mu^2)$ problem. At $O(\mu^2)$, we get
\begin{equation}
    \mathcal{H}_2(\mathcal{H}_{0,z}+\mathcal{H}_{0,zzz})+\mathcal{H}_0(\mathcal{H}_{2,z}+\mathcal{H}_{2,zzz})=\mathcal{Y}_2(\mathcal{H}_{0,z}+\mathcal{H}_{0,zzz}) -1-\mathcal{H}_1(\mathcal{H}_{1,z}+\mathcal{H}_{1,zzz}).
    \label{mu2problem}
\end{equation}
The associated boundary conditions are $\mathcal{H}_{2,z}=\mathcal{Y}_2=0$ at $z=\{0,L\}$, and mass conservation implies $\int_0^L\mathcal{H}_{2}\,\mathrm{d}z=0$. Now, if we define $\boldsymbol\varphi_2=(\mathcal{H}_2,u_2)^T$ with $u_2 = \int_0^z\mathcal{H}_2\,\mathrm{d}z$ then it satisfies a solvability condition,
\begin{equation}
    \langle\boldsymbol\phi^\dagger, \mathcal{L}\boldsymbol\varphi_2\rangle=\langle\mathcal{L}^\dagger\boldsymbol\phi^\dagger,\boldsymbol\varphi_2\rangle=0.
    \label{solvability1}
\end{equation}
Equation \eqref{mu2problem} provides an expression for $(\partial^3_z+\partial_z+\mathcal{G})\mathcal{H}_2$, 
so \eqref{solvability1} implies 
\begin{equation}
    \int_0^L\phi_1^\dagger\left(\mathcal{G}\mathcal{Y}_2-\frac{1}{\mathcal{H}_0}+\frac{\mathcal{A}^2\phi_1^2}{\mathcal{H}_0}\mathcal{G}\right)\,\mathrm{d}z=0.
    \label{solvability}
\end{equation}
The evolution equation \eqref{TFevoleqn}, at leading-order in $\mu$, can be integrated once in $z$ to give
\begin{equation}
    2\mathcal{A}_T\phi_2 = -B_*\mathcal{Y}_2^2
    \label{395}
\end{equation}
where we have used the boundary conditions $\mathcal{Y}_2=\phi_2=0$ at $z=0$. Figure \ref{fig:nearbif}(a) shows that $\phi_2\leq0$, so \eqref{395} requires $\mathcal{A}_T\geq0$. Using \eqref{395}, the solvability condition \eqref{solvability} becomes
\begin{equation}
    \int_0^L\phi_1^\dagger\left(\mathcal{G}\left[-\frac{2\mathcal{A}_T\phi_2}{B_*}\right]^{1/2}-\frac{1}{\mathcal{H}_0}+\frac{\mathcal{A}^2\phi_1^2}{\mathcal{H}_0}\mathcal{G}\right)\,\mathrm{d}z=0,
    \label{ATevoleqn}
\end{equation}
which is the evolution equation for $\mathcal{A}(T)$. Notice that \eqref{ATevoleqn} is independent of the amplitude of $\phi_1^\dagger$, so the value of $\phi_2^\dagger$ is truly arbitrary. 

After computing $\boldsymbol\phi$ and $\boldsymbol\phi^\dagger$, the coefficients in \eqref{ATevoleqn} are found using numerical integration. Equation \eqref{ATevoleqn} has two fixed points at $\mathcal{A}\approx \pm 2.20$. Figure \ref{fig:nearbif}(b) is a solution of \eqref{ATevoleqn} with initial conditions $\mathcal{A}(0)=-2.19$, showing that the solution evolves away from the negative fixed point towards the positive one, suggesting the former is unstable and the latter stable. Fixed points in $\mathcal{A}$ correspond to static solutions for $H$. Since $\phi_1(L)>0$ (figure \ref{fig:nearbif}a), the negative fixed point must correspond to the lower-branch static solution in figure \ref{fig:bifurdiagram}(a), since then $\mathcal{H}_1<0$ so $\max_zH<\max_zH_0(z;B^*)$. Similarly, the positive fixed point must correspond to the upper-branch solution since $\mathcal{H}_1>0$. This confirms that, at least near $B=B^*$, the lower-branch solutions are unstable and the upper-branch solutions are stable.

\section{Small $B$ approximation to the upper-branch thin-film static solutions}\label{section:smallB}

The upper-branch solutions in figure \ref{fig:bifurdiagram}(a) are approximated in the limit $B\rightarrow0$ using the method of matched asymptotic expansions. We identify three asymptotically distinct regions in space which can be matched together. Region I is the smaller collar which lies approximately in $0<z\lesssim L-\upi$. Region II is the thin inner region between the two collars, approximately located at $z=L-\upi$. Region III is the large collar which lies approximately in $L-\upi\lesssim z<L$. Separate asymptotic expansions for $H_0(z;B)$ will be proposed in each of the three regions. To determine the leading-order form of the expansions in each region, the following simple scaling argument is used. 

Let $L_I$, $L_{II}$, $L_{III}$ be horizontal lengthscales for Regions I, II and III, respectively. Similarly, let $H_I$, $H_{II}$, $H_{III}$ be scales for the size of $H_0$ in each of the three regions. Region I and Region III both have $O(1)$ width, so the lengthscales are $L_I\sim1$ and $L_{III}\sim1$. The width of Region II is small, $L_{II}\ll 1$; a precise scaling will be determined in the following. Since the Newtonian solution is recovered as $B\rightarrow0$, $H_0=O(1)$ in Region III so $H_{III}\sim1$. Also, the Newtonian solution has constant non-zero curvature in Region III, so $\kappa\sim 1$ in Region III. In Region II, since $L_{II}\ll 1$, the curvature is dominated by the second derivative, so $\kappa\sim H_{II}/L_{II}^2$. In order for Regions II and III to match, these curvatures must balance, so $H_{II}/L_{II}^2\sim1$. The ODE for $H_0$, \eqref{H0eqn}, can now be used to obtain the remaining scalings. In Region I, $L_I\sim1$, so \eqref{H0eqn} implies $H_I\sim B^{1/2}$. In Region II, \eqref{H0eqn} implies $H_{II}^2/L_{II}^3\sim B$. This last result, combined with $H_{II}\sim L_{II}^2$, gives $H_{II}\sim B^2$ and $L_{II}\sim B$. 

Informed by the scaling argument, we propose the following expansions for $H_0(z;B)$ in the limit $B\rightarrow0$. In Region I,
\begin{equation}
    H_0(z;B) = B^{1/2}\hat{h}_0(z) + \dots.
    \label{R1expansion}
\end{equation}
In Region II,
\begin{equation}
    H_0(z;B) = B^2\bar{h}_0(\zeta)+B^{5/2}\bar{h}_1(\zeta)+\dots, \quad\mbox{where\ }\quad z = L - \upi + B\log B\bar{z}_0 +  B\zeta + \dots\label{R2expansion}
\end{equation}
and $\bar{z}_0$ is a constant which we determine below. The $B\log B\bar{z}_0$ term in \eqref{R2expansion} determines how the location of Region II varies with $B$. In Region III,
\begin{equation}
    H_0(z;B) = \tilde{h}_0(z) + B^{1/2}\tilde{h}_1(z)  + B\tilde{h}_2(z) + \dots.
    \label{R3expansion}
\end{equation}

Inserting the expansions \eqref{R1expansion}-\eqref{R3expansion} into the ODE \eqref{H0eqn}, and equating at each order of $B$, gives
\begin{equation}
        \hat{h}_0\left(\hat{h}_{0,zzz}+\hat{h}_{0,z}\right)= 1, 
        \label{R1eqn}
\end{equation}
which holds in $0\leq z <L-\upi$,
\refstepcounter{equation}
$$
    \bar{h}_0\bar{h}_{0,\zeta\zeta\zeta} = 1, \quad \bar{h}_0^3\bar{h}_{1,\zeta\zeta\zeta}+\bar{h}_1=0, 
    \eqno{(\theequation{\mathit{a},\mathit{b}})}\label{R2eqn}
$$
which hold in $-\infty<\zeta<\infty$, and
\refstepcounter{equation}
$$
    \hat{h}_{0,zzz}+\hat{h}_{0,z}=0, \quad \hat{h}_{1,zzz}+\hat{h}_{1,z}=0,\quad \hat{h}_{2,zzz}+\hat{h}_{2,z}=\frac{1}{\tilde{h}_0}, 
    \eqno{(\theequation{\mathit{a},\mathit{b},\mathit{c}})}\label{R3eqns}
$$
which hold in $L-\upi<z\leq L$.
The boundary conditions \eqref{TFsideBCs} imply
\begin{subequations}
\begin{eqnarray}
    \hat{h}_{0,z} = 0 \quad \mbox{at\ }\quad z=0,\label{z0BCs}\\
    \tilde{h}_{0,z} = \tilde{h}_{1,z} = \tilde{h}_{2,z} = 0 \quad\mbox{at\ }\quad z=L.
    \label{zLBCs}
    \end{eqnarray}
\end{subequations}
Mass conservation implies
\refstepcounter{equation}
$$
    \int_{L-\upi}^L\tilde{h}_0\,\mathrm{d}z=L, \quad \int_{0}^{L-\upi}\hat{h}_0\,\mathrm{d}z+\int_{L-\upi}^L\tilde{h}_1\,\mathrm{d}z=0, \quad \int_{L-\upi}^L\tilde{h}_0\,\mathrm{d}z=0.
    \eqno{(\theequation{\mathit{a},\mathit{b},\mathit{c}})}\label{massBCs}
$$

The problem is closed by determining matching conditions between the regions. To match Regions II \& III, consider \eqref{R2eqn} in the limit $\zeta\rightarrow\infty$, which gives
\begin{subequations}
\begin{eqnarray}
    \bar{h}_0 &=& a_0\zeta^2-\frac{1}{a_0}\zeta\log\zeta+c_0\zeta+\frac{1}{4a_0^3}(\log\zeta)^2+\left(\frac{3}{4a_0^3}-\frac{c_0}{2a_0^2}\right)\log\zeta+\dots,
    \label{R2largezeta1}\\
    \bar{h}_1 &=& a_1\zeta^2+\frac{a_1}{a_0}\zeta\log\zeta+\dots,
    \label{R2largezeta2}
\end{eqnarray}\label{R2largezeta}
\end{subequations}
as $\zeta\rightarrow\infty$, for some constants $a_0$, $a_1$, $c_0$. To derive boundary conditions at $z=L-\upi$, we write $\zeta=(z-L+\upi)/B + \log B\bar{z}_0+\dots$ in \eqref{R2largezeta}, then equate the resulting expansion for $B^2\bar{h}_0+B^{5/2}\bar{h}_1+\dots$ with $\tilde{h}_0+B^{1/2}\tilde{h}_1+B\tilde{h}_2+\dots$ in the limit $z\rightarrow(L-\upi)^+$. 
This gives

\refstepcounter{equation}
$$
    \tilde{h}_0\sim a_0(z-L+\upi)^2, \quad
    \tilde{h}_1\sim a_1(z-L+\upi)^2 ,
    \eqno{(\theequation{\mathit{a},\mathit{b}})}\label{R3matching1}
    $$
    \begin{equation}
    \tilde{h}_2\sim (z-L+\upi)\left(c_0-\frac{1}{a_0}\log(z-L+\upi)\right),
    \label{R3matching2}
\end{equation}
as $z\rightarrow(L-\upi)^+$, and $\bar{z}_0=1/(2a_0^2)$. We find the general solutions to \eqref{R3eqns}, apply the boundary conditions \eqref{zLBCs} and mass conservation conditions \eqref{massBCs}, then expand in the limit $z\rightarrow(L-\upi)^+$ and match this to \eqref{R3matching1}-\eqref{R3matching2}. This gives $a_0=L/(2\upi)$, $c_0=2\upi\log 2/L$, 
\refstepcounter{equation}
$$
\tilde{h}_0 = \frac{L}{\upi}\left[1+\cos(z-L)\right],\quad
    \tilde{h}_1 = 2a_1\left[1+\cos(z-L)\right],
\eqno{(\theequation{\mathit{a},\mathit{b}})}
\label{R3soln}
$$
\begin{eqnarray}
    \tilde{h}_2 &=&  \frac{2}{L} + \frac{\upi}{L}\sin(z-L) + \frac{2-\upi^2}{L}\cos(z-L) \nonumber\\&&\mbox{} +\frac{\upi}{L}\left[(L-z)\cos(z-L) + 2\sin(z-L)\log\left(\cos\left(\frac{z-L}{2}\right)\right)\right],\label{R3soln}
\end{eqnarray}
and $2\upi a_1=-\int_0^{L-\upi}\hat{h}_0\,\mathrm{d}z$, which is determined numerically once $\hat{h}_0$ is found. 

To determine matching conditions between Regions I \& II, we follow a similar process. Expanding now in the limit $\zeta\rightarrow-\infty$ gives
\begin{equation}
    \bar{h}_0\sim \sqrt{\frac{8}{3}}(-\zeta)^{3/2}+\dots \quad\mbox{as\ }\quad \zeta\rightarrow-\infty,
    \label{largenegzeta}
\end{equation}
from which we infer the boundary condition
\begin{equation}
    \hat{h}_0\sim \sqrt{\frac{8}{3}}(L-\upi-z)^{3/2} \quad\mbox{as\ }\quad  z\rightarrow(L-\upi)^-.
    \label{R1matching}
\end{equation}
The Region I problem is equation \eqref{R1eqn} subject to boundary conditions \eqref{z0BCs} and \eqref{R1matching}. When solving the Region I problem, we define a small constant $\hat\epsilon$ and solve \eqref{R1eqn} in the domain $0\leq z\leq L-\upi-\hat\epsilon$. We enforce \eqref{R1matching} by setting $\hat{h}_0(L-\upi-\hat\epsilon)=\sqrt{8/3}\hat\epsilon^{3/2}$ and $\hat{h}_0'(L-\upi-\hat\epsilon)=\sqrt{6}\hat\epsilon^{1/2}$. We choose $\hat\epsilon$ sufficiently small that the solution in the rest of the domain is insensitive to its exact value. 

\begin{figure}
    \centering
    \includegraphics[width=\textwidth]{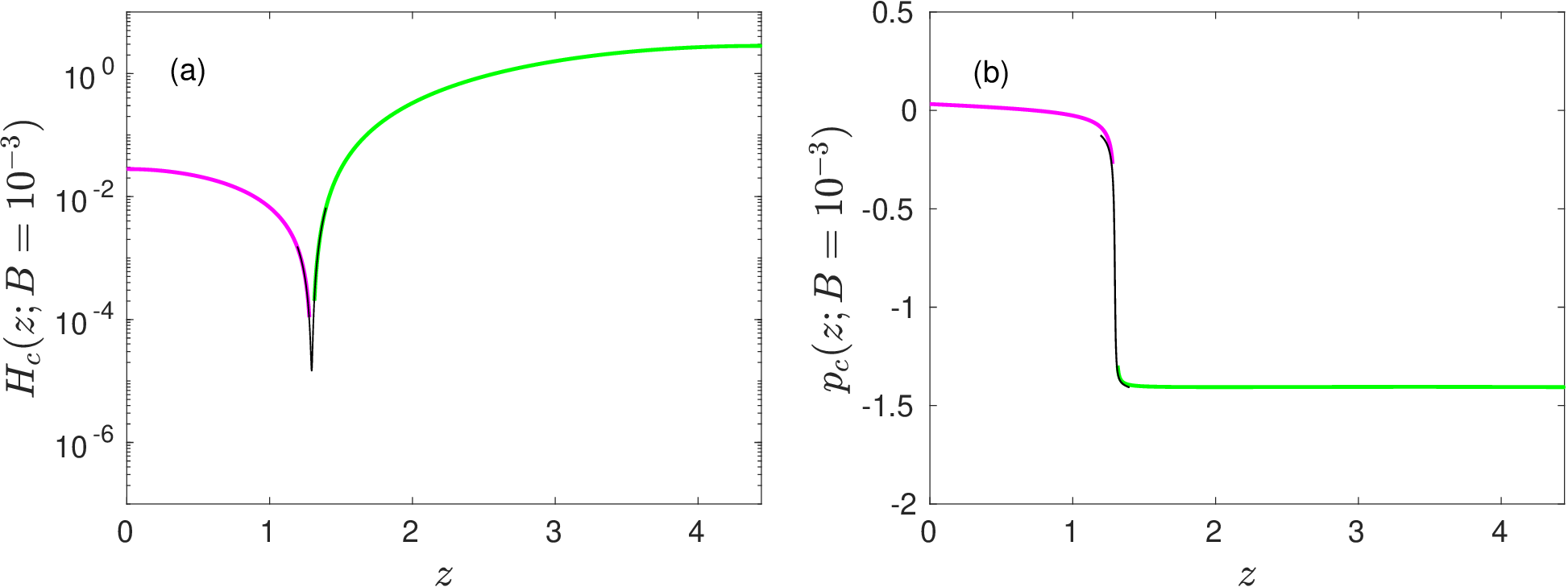}
    \caption{Approximation to the upper-branch static solution (figure \ref{fig:bifurdiagram}a) with $B=10^{-3}$ using matched asymptotic expansions. (a) Composite solution, $H_c(z;10^{-3})$, which is composed of the solutions in Regions I (magenta), II (black) and III (green). (b) Capillary pressure of the composite solution, $p_c\equiv -H_c-H_{c,zz}$.}
    \label{fig:smallBMAEsoln}
\end{figure}

When solving the Region II problem to find $\bar{h}_0$, we define a large constant $\zeta_\infty$ and solve (\ref{R2eqn}\textit{a}) in the finite domain $-\zeta_\infty<\zeta<\zeta_\infty$. Informed by \eqref{R2largezeta1} and \eqref{largenegzeta}, we enforce the boundary conditions,
\begin{equation}
    \bar{h}_{0,\zeta\zeta} = \frac{\sqrt{6}}{2}\zeta_\infty^{-1/2} \quad\mbox{at\ }\quad\zeta=-\zeta_\infty,
\end{equation}
\refstepcounter{equation}
$$
    \bar{h}_0=\frac{L}{2\upi}\zeta_\infty^2-\frac{2\upi}{L}\zeta_\infty\log\zeta_\infty, \quad \bar{h}_{0,\zeta\zeta} = \frac{L}{\upi}+\frac{\upi}{L\zeta_\infty} \quad\mbox{at\ }\quad\zeta=\zeta_\infty.\eqno{(\theequation{\mathit{a},\mathit{b}})}
$$
We choose $\zeta_\infty$ large enough that the solution is insensitive to its exact value away from the boundaries $\zeta=\pm\zeta_\infty$.

Figure \ref{fig:smallBMAEsoln}(a) shows the composite solution, which we call $H_c(z;B)$, for $B=10^{-3}$. To compute this solution, we used $\hat\epsilon=10^{-4}$ and $\zeta_\infty=600$. The value $a_1\approx-0.105$ computed with this solution completes the Region III solution \eqref{R3soln}. The constant $a_1$ depends on $L$, which here is $L=\sqrt{2}\upi$. For clarity, the solutions displayed in figure \ref{fig:smallBMAEsoln} are truncated shortly after the points where they overlap. This also means that the parts of the solutions displayed are away from the boundaries so entirely independent of the values of $\hat\epsilon$ and $\zeta_\infty$ chosen. Figure \ref{fig:smallBMAEsoln}(b) shows the capillary pressure, $p_{c}\equiv-H_{c}-H_{c,zz}$, of the composite solution.

\section{Static solutions of the long-wave evolution equation}\label{section:LWstatics}

Following our approach in §\ref{section:TFstatics}, we look for marginally-yielded static solutions, $R=R_0(z;B,\epsilon)$, of the long-wave equations \eqref{psidefn}-\eqref{LWsideBCs2}. The static shapes $R_0(z;B,\epsilon)$ are solutions to $\psi=1$ where $\psi$ is defined in \eqref{psidefn}. As in the thin-film analysis, we assume that the pressure is monotonic, so $\hat{p}_z<0$. The ODE $\psi=1$ can be rearranged to give
\begin{equation}
    \hat p_z(1-R_0^2)=-2\epsilon^2B,\label{LWpsi1}
\end{equation}
where $\hat{p}$ is defined in \eqref{LWpressuregrad} and $\epsilon^2B=\hat{B}$. 
We solve \eqref{LWpsi1} subject to boundary conditions $R_{0,z}(0;B,\epsilon)=R_{0,z}(L;B,\epsilon)=0$, and the volume conservation condition,
\begin{equation}
    2\upi\int_0^L(1-R_0^2)\,\mathrm{d}z=2\upi\epsilon L\left(2-\epsilon\right).
\end{equation}
The problem is solved using a boundary value problem solver in \textsc{Matlab}. We define the layer thickness, $H_0(z;B,\epsilon)\equiv(1-R_0)/\epsilon$, to aid discussion and comparison with the thin-film static solutions computed in §\ref{section:TFstatics}. 

\begin{figure}
    \centerline{\includegraphics[trim=0 0 0 -8mm,width=\textwidth]{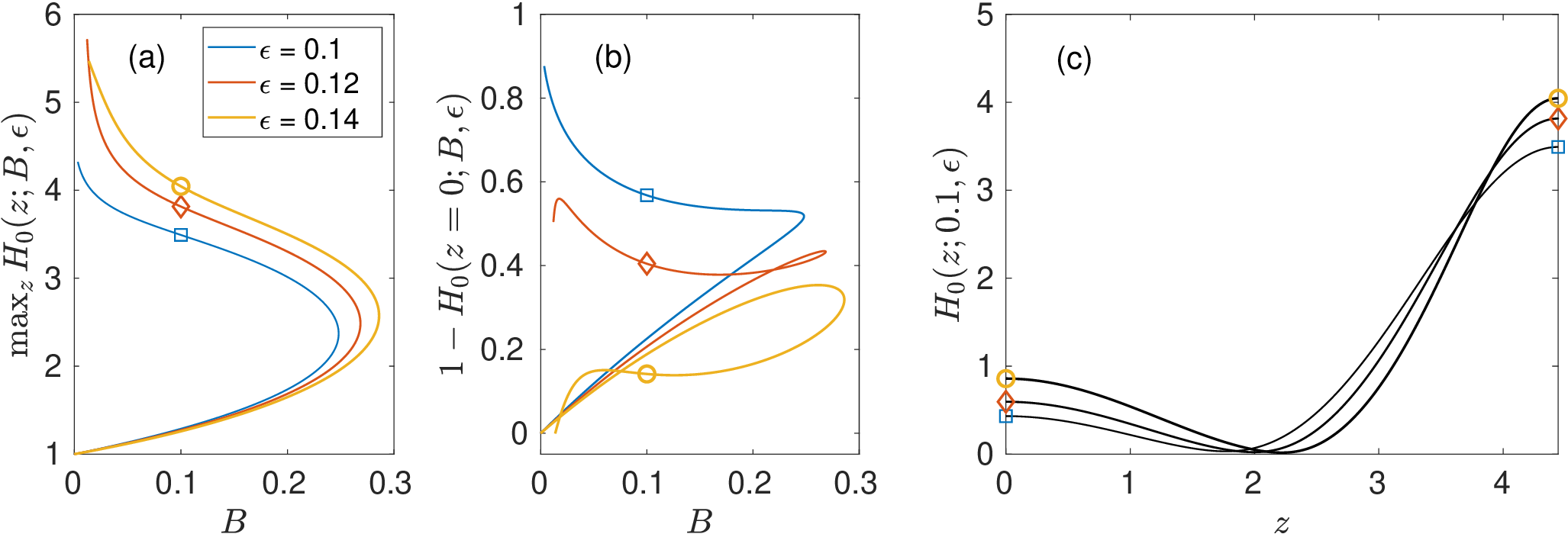}}
    \caption{Static solutions, $H_0(z;B,\epsilon)$, satisfying equation \eqref{LWpsi1} with monotonic curvature, for $\epsilon=0.1, 0.12, 0.14$. (a) Maximum height of solutions, $\max_zH_0(z;B,\epsilon)$, (b) $1-H_0(z=0;B,\epsilon)$, and (c) three example upper-branch solutions at $B=0.1$, which correspond to the locations indicated by the markers on (a) and (b). 
    }
    \label{fig:LWstatics}
\end{figure}

Figure \ref{fig:LWstatics} shows solutions for $\epsilon=0.1,0.12,0.14$. Figure \ref{fig:LWstatics}(a) shows that, like in the thin-film case, we find an upper and a lower branch of solutions for each $\epsilon$, and a bifurcation point $B=B^*_\epsilon$ such that no solutions exist for $B>B^*_\epsilon$. Note that the location of the bifurcation now depends on $\epsilon$. The boundary value problem solver is generally able to compute the whole lower branch and most of the upper branch of solutions, except for very small $B$. The upper-branch solutions are very singular for small $B$, with an increasingly large jump in $H_{0,zz}$ around the minimum in $H_0$, which makes computing them difficult.

Figure \ref{fig:LWstatics}(b) shows plots of $1-H_0(0;B,\epsilon)$. In §\ref{section:dependenceTF}, we show that the same quantity from the thin-film static solutions has particular significance in determining the outcome of an evolving thin layer. Figure \ref{fig:LWAvsB} shows that it has much less physical significance in the long-wave problem when $\epsilon\geq0.12$. We argue that this is because the evolving layer generally does not select a static shape which is a solution to \eqref{LWpsi1}. Instead, either a plug forms, or the layer rigidifies near $z=0$ early in the evolution which leads to a different static two-collar solution being selected. 

\bibliographystyle{jfm}
\bibliography{evolution}

\end{document}